\documentclass[aps,prd,amsfonts,amssymb,amsmath,twocolumn,showpacs,eqsecnum,floatfix]{revtex4}

\usepackage{graphicx}
\usepackage{bm}

\newcommand{\bmath}[1]{\mbox{{\boldmath{{$#1$}}}}}

\begin{document}

\title{Vacuum polarization for lukewarm black holes}

\author{Elizabeth Winstanley} \email{E.Winstanley@sheffield.ac.uk}
\affiliation{Department of Applied Mathematics, The University of
Sheffield, Hicks Building, Hounsfield Road, Sheffield, S3 7RH,
United Kingdom.}
\author{Phil M. Young}
\affiliation{Department of Applied Mathematics, The University of
Sheffield, Hicks Building, Hounsfield Road, Sheffield, S3 7RH,
United Kingdom.}

\date{\today }

\begin{abstract}
We compute the renormalized expectation value of the square of a quantum scalar field on
a Reissner-Nordstr\"om-de Sitter black hole in which the temperatures of the event and
cosmological horizons are equal (`lukewarm' black hole).
Our numerical calculations for a thermal state at the same temperature as the two horizons
indicate that this renormalized expectation value is regular on both the event and cosmological
horizons.
We are able to show analytically, using an approximation for the
field modes near the horizons, that this is indeed the case.
\end{abstract}


\maketitle

\section{Introduction}
\label{sec:intro}

The renormalized stress-energy tensor (RSET) $\langle T_{\mu \nu } \rangle _{\rm {ren}}$
is an object of fundamental
importance in quantum field theory in curved space-time, since it governs, via
the semi-classical Einstein equations
\begin{equation*}
G_{\mu \nu } = 8\pi G \langle T_{\mu \nu } \rangle _{\rm {ren}},
\end{equation*}
the back-reaction of the quantum field on the space-time geometry.
Computing the RSET on a particular space-time background is a complicated process \cite{ash,Tmunucalcs,howard,jo},
and, for a quantum
scalar field $\phi $, it is informative to study first the renormalized vacuum polarization
$\langle \phi ^{2} \rangle _{\rm {ren}}$, which is considerably easier to compute and has many of the
same features as the full RSET.
For example, although $\langle \phi ^{2} \rangle _{\rm {ren}}$ is a scalar object and hence cannot distinguish between
future and past event horizons, nonetheless, if it diverges at a horizon for a particular quantum state, then
it is likely that the RSET also diverges there.

The vacuum polarization $\langle \phi ^{2} \rangle _{\rm {ren}}$ has been extensively studied by many authors
for various black hole backgrounds, beginning with computations by Candelas \cite{candelas} of
$\langle \phi ^{2} \rangle _{\rm {ren}}$ for a massless, conformally coupled, scalar field
on the event horizon (and at infinity) for a Schwarzschild black hole.
This was subsequently extended to the whole of the Schwarzschild geometry in Refs. \cite{ch} (exterior to the
event horizon) and \cite{cj} (interior to the event horizon).
The corresponding calculation for massive scalar fields was done by Anderson \cite{andersonm}.
Other examples of
calculations of $\langle \phi ^{2} \rangle _{\rm {ren}}$ on more general black hole space-times can be found
in \cite{phi2calcs}.
As well as exact, numerical calculations, a number of approximation schemes have been developed for various
types of space-times, both for $\langle \phi ^{2} \rangle _{\rm {ren}}$ \cite{phi2approx} and the RSET
\cite{Tmunuapprox}.

Of the three standard vacua for quantum fields on black hole space-times
(Hartle-Hawking \cite{hh}, Unruh \cite{u} or Boulware \cite{b}), the Hartle-Hawking state has received
the most attention in the literature.
This is because the Hartle-Hawking vacuum possesses the most symmetries (for example, time reversal symmetry)
and so is more straightforward to calculate.
Expectation values of observables in this state are also expected to be regular on the event horizon of
a black hole.
Once the renormalized expectation values in a particular state (say the Hartle-Hawking state) have been
computed, it is much easier to compute renormalized expectation values in another state because
the difference between expectation values in two states does not require renormalization.
A general method for computing $\langle \phi ^{2} \rangle _{\rm {ren}}$ on any static, spherically symmetric,
black hole space-time for a state at either a fixed non-zero temperature or zero temperature
was developed by Anderson \cite{anderson}, and subsequently extended to calculations
of the RSET \cite{ash} (see also \cite{satz} for an application of this type of approach to the space-time outside
a star).
This is the method we shall adopt in this paper.
This approach works for computations of the Hartle-Hawking (non-zero temperature) and Boulware (zero temperature)
states, but not for the Unruh vacuum.

Most of the work on $\langle \phi ^{2} \rangle _{\rm {ren}}$ and the RSET to date has focussed on
asymptotically flat black holes, such as Schwarzschild and Reissner-Nordstr\"om.
However, there are many interesting features of quantum field theory on asymptotically de Sitter  black holes
\cite{sds}, when we have both a black hole event horizon and a cosmological horizon.
The Kay-Wald theorem \cite{kay} states that there is no thermal state on Schwarzschild-de Sitter black holes
which preserves all the symmetries of the metric (including time-reversal symmetry) and which is regular
on both the event and cosmological horizons.
Therefore there is no equivalent of the Hartle-Hawking state for this geometry.
This may be understood heuristically as follows.
A thermal state is expected to be regular at a horizon if the temperature of the state matches the temperature
of the horizon.
However, for Schwarzschild-de Sitter black holes, the temperatures of the event and cosmological horizons
are not equal and therefore any thermal state cannot match both temperatures.
If we consider the more general case of Reissner-Nordstr\"om-de Sitter black holes, then it is possible
for the event and cosmological horizons to have the same temperature \cite{mellor}
(these are known as `lukewarm' black holes \cite{romans}).
Therefore a natural question is whether the thermal state, constructed on these black holes at the natural
temperature, is regular on both the event and cosmological horizons.
A simple calculation for two-dimensional black holes \cite{mtwy} shows that this is the case,
but for four-dimensional black holes there is an unknown function in the RSET which can only be found by
direct calculation.
We will return to the computation of this function in the near future \cite{owy}, but in the present
paper we will focus on the simpler computation of $\langle \phi ^{2} \rangle _{\rm {ren}}$ for this
state.

The outline of this paper is as follows.
In section \ref{sec:ash} we review the method of \cite{ash} to calculate $\langle \phi ^{2} \rangle _{\rm {ren}}$
using point-splitting.
As we are considering a thermal quantum state, the Euclidean Green's function will be used.
Using temporal point-splitting, even with the points split the Green's function contains apparent divergences.
We present in section \ref{sec:finite} a new approach to regularizing these divergences, based on dimensional reduction.
Once the point-split Green's function is manifestly finite, the renormalization procedure is then relatively
straightforward, following \cite{ash}.
The renormalized expectation value $\langle \phi ^{2} \rangle _{\rm {ren}}$ is computed numerically
for lukewarm black holes in section \ref{sec:numerics}.
Our numerical results indicate that $\langle \phi ^{2} \rangle _{\rm {ren}}$ is indeed regular on both the
event and cosmological horizons.
This is proved analytically in section \ref{sec:regular}, in which we use an approximation
for the field modes near the horizons, developed in Ref. \cite{tomimatsu}
for the asymptotically flat Reissner-Nordstr\"om case.
Finally, our conclusions are presented in section \ref{sec:conc}.
Throughout this paper the (Lorentzian) metric has signature $(-,+,+,+)$ and we use units in which
$8\pi G = \hbar = c = k_{B} = 1$.

\section{General Method to Calculate $\langle \phi ^{2} \rangle _{\rm {ren}}$}
\label{sec:ash}

\subsection{Point-split Green's Function}
\label{sec:unren}

We use the method of Ref. \cite{anderson,ash} to compute the renormalized expectation value
$\langle \phi ^{2} \rangle _{\rm {ren}}$.
In this section we just briefly outline the key steps in the construction, further details
of which can be found in Refs. \cite{ash,anderson}.

We begin with a scalar field $\phi $
with mass $m$ and coupling $\xi $ to the scalar curvature $R$, satisfying the field equation
\begin{equation}
\left( \nabla _{\mu } \nabla ^{\mu } - m^{2} - \xi R \right) \phi = 0.
\end{equation}
In this paper we consider a black hole space-time with metric, in Schwarzschild-like co-ordinates, given by
\begin{equation}
ds^{2} = - f(r) \, dt^{2} + f(r) ^{-1} \, dr^{2} + r^{2} \, d\theta ^{2}
+ r^{2} \sin ^{2} \theta \, d\varphi ^{2},
\label{eq:metric}
\end{equation}
where the metric function $f$ depends only on the radial co-ordinate $r$.
The metric (\ref{eq:metric}) is not the most general spherically symmetric black hole metric, and,
indeed, the method of Ref. \cite{ash} is developed for more general spherically symmetric metrics.
However, the metric (\ref{eq:metric}) is sufficiently general to cover a broad class of black hole
geometries, including, in particular, the Reissner-Nordstr\"om-de Sitter black holes for which
\begin{equation}
f(r)=1-\frac {2M}{r}+ \frac {Q^{2}}{r^{2}} -\frac {\Lambda r^{2}}{3},
\label{eq:fRNdS}
\end{equation}
where $M$, $Q$ are related, respectively, to the mass and charge of the black hole \cite{eugen}
and $\Lambda $ is the (positive) cosmological constant.
Further details of the space-time geometry in which we are particularly interested, the
lukewarm black holes, will be given in section \ref{sec:lukewarm}.

We are interested in computing $\langle \phi ^{2} \rangle _{\rm {ren}}$
for a thermal state at a temperature $T$.
We follow \cite{ash} and use a Euclidean space approach.
By defining the Euclidean time $\tau $ as $\tau = it$, the metric (\ref{eq:metric})
becomes
\begin{equation}
ds^{2} = f(r) \, d\tau ^{2} + f(r) ^{-1} \, dr^{2} + r^{2} \, d\theta ^{2}
+ r^{2} \sin ^{2} \theta \, d\varphi ^{2}.
\label{eq:Emetric}
\end{equation}
We define $G_{E}(x;x')$ to be the Euclidean Green's function, which satisfies the equation \cite{ash}
\begin{equation}
\left[ \nabla _{x\mu } \nabla _{x}^{\mu } - m^{2} - \xi R \right] G_{E}(x;x') =
- g^{-\frac {1}{2}}(x) \delta ^{4} (x,x'),
\label{eq:GEDE}
\end{equation}
where the covariant derivative is now taken with respect to the metric (\ref{eq:Emetric}).
Using the method of point-splitting, the (unrenormalized) expectation value
$\langle \phi ^{2} \rangle _{\rm {unren}}$ is given by the following limit:
\begin{equation}
\langle \phi ^{2} \rangle _{\rm {unren}} = \Re \left[ \lim _{x\rightarrow x'} G_{E}(x;x') \right] .
\end{equation}
The stress-energy tensor is calculated from derivatives of $G_{E}(x;x')$ \cite{ash}.

For a thermal state at a temperature $T$, the Euclidean Green's function $G_{E}(x;x')$ is periodic
in $\tau - \tau' $ with period $T^{-1}$.
In this case the scalar Euclidean Green's function takes the form \cite{ash,anderson}
\begin{eqnarray}
G_{E}(x;x')  & = &
\frac {T}{4\pi } \sum _{n=-\infty }^{\infty } \exp \left[ i \omega \left( \tau - \tau ' \right) \right]
\nonumber \\ & & \times
\sum _{\ell = 0}^{\infty } \left( 2 \ell + 1 \right) P_{\ell }(\cos \gamma ) \Upsilon _{\omega \ell } (r,r'),
\label{eq:split}
\end{eqnarray}
where $\omega = 2\pi n T$, and $P_{\ell }$ is the usual Legendre function, with
\begin{equation}
\cos \gamma = \cos \theta \cos \theta ' + \sin \theta \sin \theta ' \cos \left( \varphi - \varphi ' \right) ,
\label{eq:gamma}
\end{equation}
and $\Upsilon _{\omega \ell }$ satisfies the differential equation
\begin{eqnarray}
 -\frac {1}{r^{2}} \delta \left( r-r' \right) & = & f \frac {d^{2} \Upsilon _{\omega \ell }}{dr^{2}}
+ \left( \frac {2f}{r} + \frac {df}{dr} \right) \frac {d\Upsilon _{\omega \ell }}{dr}
\nonumber
 \\ & &
- \left[ \frac {\omega ^{2}}{f} + \frac {\ell (\ell + 1)}{r^{2}} + m^{2} + \xi R \right] \Upsilon _{\omega \ell } .
\nonumber \\
\label{eq:UpsilonDE}
\end{eqnarray}
The differential equation (\ref{eq:UpsilonDE}) arises from separating the wave equation (\ref{eq:GEDE}) on the background
metric (\ref{eq:Emetric}).

We define functions $p_{\omega \ell }$ and $q_{\omega \ell }$ as solutions of the corresponding homogeneous
differential equation
\begin{eqnarray}
 0 & = & f \frac {d^{2} \Upsilon _{\omega \ell }}{dr^{2}}
+ \left( \frac {2f}{r} + \frac {df}{dr} \right) \frac {d\Upsilon _{\omega \ell }}{dr}
\nonumber
 \\ & &
- \left[ \frac {\omega ^{2}}{f} + \frac {\ell (\ell + 1)}{r^{2}} + m^{2} + \xi R \right] \Upsilon _{\omega \ell } ,
%\nonumber \\
\label{eq:modes}
\end{eqnarray}
with appropriate boundary conditions.
These will be discussed further in section \ref{sec:mode} for the
particular case of lukewarm black holes.
Typically, $p_{\omega \ell }$ is the solution which is regular at the lower limit of the region under consideration
(for black holes, the event horizon), while $q_{\omega \ell }$ is regular at the upper limit of the region
(usually infinity, but in our case the cosmological horizon).
The function $\Upsilon _{\omega \ell }(r,r')$ is then given by \cite{anderson}
\begin{equation}
\Upsilon _{\omega \ell } (r,r') = C_{\omega \ell } p_{\omega \ell }(r_{<}) q_{\omega \ell }(r_{>}),
\label{eq:Upsilondef}
\end{equation}
where, as usual, $r_{<}$ is the lesser of the two values $r$, $r'$ and $r_{>}$ is the greater.
In (\ref{eq:Upsilondef}), the normalization constant $C_{\omega \ell }$ is fixed by the normalization condition
\cite{anderson}
\begin{equation}
C_{\omega \ell } \left[ p_{\omega \ell } \frac {dq_{\omega \ell }}{dr}
- q_{\omega \ell } \frac {dp_{\omega \ell }}{dr}  \right]
= - \frac {1}{r^{2}f}.
\label{eq:Wronskian}
\end{equation}
Further properties of the mode functions $p_{\omega \ell }$ and $q_{\omega \ell }$ will be discussed in section
\ref{sec:mode}.

We now follow standard procedure and choose temporal point-splitting, so that $r=r'$, $\theta = \theta '$
and $\varphi = \varphi '$.
Then $\gamma =1$ (\ref{eq:gamma}) and, using $P_{\ell }(1)=1$, we have
\begin{eqnarray}
G_{E}(\tau , {\bmath {x}}; \tau ', {\bmath {x}} ) & = &
\frac {T}{4\pi } \sum _{n=-\infty }^{\infty }e^{i \omega \epsilon }
\nonumber
 \\ & & \times
\sum _{\ell = 0}^{\infty } \left( 2 \ell + 1 \right) C_{\omega \ell } p_{\omega \ell }(r) q_{\omega \ell }(r),
\nonumber \\
\label{eq:splitinfinite}
\end{eqnarray}
where $\epsilon = \tau - \tau ' $.

\subsection{Finite Mode Sums}
\label{sec:finite}

Even though the points are separated, the Green's function (\ref{eq:splitinfinite}) is apparently divergent.
This is due to our choice of point-splitting and arises in practice because the sums over $\ell $ in
(\ref{eq:splitinfinite}) do not converge.
This problem is well-known, arising first in Candelas' \cite{candelas} calculations on a Schwarzschild background.
This divergence cannot be `real' since, by definition, the Green's function {\em {must}} be finite
when the points are separated.
The apparent divergences are removed by subtracting from (\ref{eq:splitinfinite}) a suitable multiple of the
delta function (which vanishes when the points are separated).
The answer is well-known, and given by \cite{ash,anderson}:
\begin{eqnarray}
G_{E}(\tau , {\bmath {x}}; \tau ', {\bmath {x}} ) & = &
\frac {T}{4\pi } \sum _{n=-\infty }^{\infty }e^{i \omega \epsilon }
\nonumber  \\ & & \hspace{-1cm} \times
\sum _{\ell = 0}^{\infty } \left[ \left( 2 \ell + 1 \right) C_{\omega \ell } p_{\omega \ell }(r) q_{\omega \ell }(r)
- \frac {1}{rf^{\frac {1}{2}}} \right] .
\nonumber \\
\label{eq:splitfinite}
\end{eqnarray}
Although this answer has been previously calculated, in this section we would like to take a different approach
to deriving the result (\ref{eq:splitfinite}) based on dimensional reduction.
This new method may prove to be useful in more complicated situations (such as Kerr black holes).

We begin by writing the general Euclidean Green's function (\ref{eq:split}) in the form
\begin{eqnarray}
G_{E}(x;x')  &  = &
 \frac {T}{4\pi } \sum _{n=-\infty }^{\infty } \exp \left[ i \omega \left( \tau - \tau ' \right) \right]
 \nonumber \\ & & \times
{\cal {G}}_{\omega } \left( r, \theta, \varphi ; r' , \theta ', \varphi ' \right) ,
\end{eqnarray}
where
\begin{equation}
{\cal {G}}_{\omega }\left( r,\theta , \varphi ; r',\theta ' ,\varphi ' \right)  =
\sum _{\ell = 0}^{\infty } \left( 2 \ell + 1 \right) \Upsilon _{\omega \ell } (r,r') P_{\ell }(\cos \gamma ).
\label{eq:3dGsplit}
\end{equation}
Using the wave equation (\ref{eq:GEDE}), the function
${\cal {G}}_{\omega }\left( r,\theta , \varphi ; r',\theta ' ,\varphi ' \right)$ satisfies the differential equation
\begin{eqnarray}
-\frac {\delta ({\bmath {x}},{\bmath {x}}')}{r^{2} \sin \theta } & = &
\frac {1}{r^{2}} \frac {\partial }{\partial r}\left[ fr^{2} \frac {\partial {\cal {G}}_{\omega }}{\partial r}
\right]
+ \frac {1}{r^{2}\sin \theta } \frac {\partial }{\partial \theta } \left[
\sin \theta \frac {\partial {\cal {G}}_{\omega }}{\partial \theta } \right]
\nonumber \\ & &
+ \frac {1}{r^{2}\sin ^{2} \theta } \frac {\partial ^{2}{\cal {G}}_{\omega }}{\partial \varphi ^{2}}
- \left[ \frac {\omega ^{2}}{f} + m^{2} +\xi R \right] {\cal {G}}_{\omega }.
\nonumber \\
\label{eq:3done}
\end{eqnarray}
Eq. (\ref{eq:3done}) looks very much like a wave equation in three dimensions, with a potential term:
\begin{equation}
{\tilde {V}} = \frac {\omega ^{2}}{f} + m^{2} +\xi R.
\end{equation}
In fact, by multiplying (\ref{eq:3done}) by $f^{-1}$, we obtain the wave equation on the three-metric
\begin{equation}
d{\tilde {s}}^{2} = dr ^{2} + r^{2} f \, d\theta ^{2} + r^{2} f \sin ^{2} \theta \, d\varphi ^{2},
\label{eq:3metric}
\end{equation}
namely:
\begin{equation}
\left[ {\tilde {\nabla }}_{i} {\tilde {\nabla }}^{i} - V({\bmath {x}}) \right]
{\cal {G}}_{\omega }({\bmath {x}};{\bmath {x}}') =
- {\tilde {g}}^{-\frac {1}{2}}({\bmath {x}}) \delta ^{3} ({\bmath {x}},{\bmath {x}}'),
\label{eq:3DGEDE}
\end{equation}
where the covariant derivatives are with respect to the three-metric (\ref{eq:3metric}) (acting on ${\bmath {x}}$)
and ${\tilde {g}}$ is the determinant of the three-metric (\ref{eq:3metric}), with the potential
\begin{equation}
V({\bmath {x}})=\frac {{\tilde {V}}}{f} = \frac {\omega }{f^{2}}  + \frac {m^{2}+\xi R}{f},
\label{eq:3Dpot}
\end{equation}
where it should be stressed that $R$ is the Ricci scalar of the {\em {original}}, four-dimensional,
metric (\ref{eq:Emetric}).

The three-metric (\ref{eq:3metric}) is curious.
We emphasize that it has no physical significance, and in
fact it has a curvature singularity at an event horizon where $f$ vanishes.
Furthermore, the potential (\ref{eq:3Dpot}) is also divergent at a horizon.
On the other hand, if we are dealing with an asymptotically flat metric (\ref{eq:metric}), then the three-metric
(\ref{eq:3metric}) is also asymptotically flat as $f\rightarrow 1$ at infinity.
The three-metric should simply be regarded as a useful mathematical tool.

From Eq. (\ref{eq:3DGEDE}), we can see that ${\cal {G}}_{\omega }({\bmath {x}};{\bmath {x}}')$ is indeed
a three-dimensional Euclidean Green's function for a scalar field on the three-metric (\ref{eq:3metric}), but
with an unusual potential (\ref{eq:3Dpot}).
However, the potential does not affect general form of
the singularity structure of the Green's function, which has the usual
Hadamard form in three dimensions \cite{friedlander,folacci}:
\begin{equation}
{\cal {G}}_{\omega } ({\bmath {x}},{\bmath {x}}')
= \frac {U({\bmath {x}},{\bmath {x}}')}{[2\sigma ({\bmath {x}},{\bmath {x}}')]^{\frac {1}{2}}}
+ W({\bmath {x}},{\bmath {x}}').
\label{eq:3DHadamard}
\end{equation}
Here, $2\sigma ({\bmath {x}},{\bmath {x}}')$ is the square of the geodesic distance between two closely
separated points ${\bmath {x}}$, ${\bmath {x}}'$, and $U({\bmath {x}}, {\bmath {x}}')$, $W({\bmath {x}},{\bmath {x}}')$
are symmetric biscalars which are regular in the limit ${\bmath {x}} \rightarrow {\bmath {x}}'$,
whose precise form will depend on the potential $V({\bmath {x}})$.
Note that there is no logarithmic term in the Hadamard expansion (\ref{eq:3DHadamard}) as we are currently
working in three rather than four dimensions.
The biscalars $U({\bmath {x}}, {\bmath {x}}')$ and $W({\bmath {x}},{\bmath {x}}')$ can be expanded in terms of
$\sigma ({\bmath {x}},{\bmath {x}}')$ using standard methods \cite{folacci,poisson}.
For our purposes here, we only require the lowest order term in $U({\bmath {x}}, {\bmath {x}}')$:
\begin{equation}
U({\bmath {x}}, {\bmath {x}}') = 1 + O(\sigma ),
\end{equation}
which does not depend on the potential $V$ (\ref{eq:3Dpot}).

We now choose a point-splitting for the three-dimensional Green's function ${\cal {G}}_{\omega }$.
We choose $r=r'$ and $\varphi = \varphi '$.
Since the metric (\ref{eq:3metric}) is spherically symmetric, we can, without loss of generality, fix
$\theta ' = 0$, and then $\cos \gamma = \cos \theta $ (\ref{eq:gamma}) and (\ref{eq:3dGsplit}) takes the form
\begin{eqnarray}
{\cal {G}}_{\omega }\left( r,\theta , \varphi ; r,0 ,\varphi  \right)
& & \nonumber \\ & & \hspace{-2.5cm}
 =
\sum _{\ell = 0}^{\infty } \left( 2 \ell + 1 \right) C_{\omega \ell }p_{\omega \ell } (r)
q_{\omega \ell }(r) P_{\ell }(\cos \theta  ).
\end{eqnarray}
Although we have brought the radial co-ordinates together, this sum is still finite because of the
$P_{\ell }(\cos \theta )$ terms.
For this point splitting, to leading order we have
\begin{equation}
2\sigma = r^{2} f \theta ^{2} + O(\theta ^{4}).
\end{equation}
Therefore the Hadamard form (\ref{eq:3DHadamard}) reads
\begin{equation}
{\cal {G}}_{\omega }\left( r,\theta , \varphi ; r,0 ,\varphi  \right)  =
\frac {1}{rf^{\frac {1}{2}}\theta } + {\mbox {finite terms}},
\end{equation}
so that, for small $\theta $, we have
\begin{equation}
\sum _{\ell = 0}^{\infty } \left( 2 \ell + 1 \right) C_{\omega \ell }p_{\omega \ell } (r)
q_{\omega \ell }(r) P_{\ell }(\cos \theta  )
= \frac {1}{rf^{\frac {1}{2}}\theta } + O(1).
\end{equation}
This clearly shows that the sums over $\ell $ in (\ref{eq:splitinfinite}) diverge.

We find the appropriate subtraction term to render the sums over $\ell $ finite using the identity \cite{howard}:
\begin{equation}
\sum _{\ell =0}^{\infty } P_{\ell }(\cos \theta ) = \frac {1}{\theta } + O (\theta ).
\end{equation}
Multiplying this by $1/(rf^{1/2})$ and subtracting, we therefore obtain:
\begin{equation}
\sum _{\ell = 0}^{\infty }\left[ \left( 2 \ell + 1 \right) C_{\omega \ell }p_{\omega \ell } (r)
q_{\omega \ell }(r) - \frac {1}{rf^{\frac {1}{2}}} \right]  P_{\ell }(\cos \theta  )
= O(1) .
\end{equation}
We may now take the limit $\theta \rightarrow 0$ and find that the sums over $\ell $
in (\ref{eq:splitfinite}) are  finite as required.

\begin{widetext}

\subsection{Renormalized Expectation Value}
\label{sec:renormalized}

From (\ref{eq:splitfinite}), we now have an expression for the unrenormalized expectation value
$\langle \phi ^{2} \rangle _{\rm {unren}}$:
\begin{eqnarray}
\langle \phi ^{2} \rangle _{\rm {unren}}
 & = & \lim _{\epsilon \rightarrow 0} \left\{ \frac {T}{4\pi } \sum _{n=-\infty }^{\infty }\cos ( \omega \epsilon )
%\right. \nonumber \\ & & \hspace{-1cm} \left. \times
\sum _{\ell = 0}^{\infty } \left[ \left( 2 \ell + 1 \right) C_{\omega \ell } p_{\omega \ell }(r) q_{\omega \ell }(r)
- \frac {1}{rf^{\frac {1}{2}}} \right]
\right\}
\nonumber \\
& = &
\lim _{\epsilon \rightarrow 0} \left\{ \frac {T}{2\pi } \sum _{n=1 }^{\infty }\cos ( \omega \epsilon )
%\right. \nonumber \\ & & \hspace{-1cm} \times
\sum _{\ell = 0}^{\infty } \left[ \left( 2 \ell + 1 \right) C_{\omega \ell } p_{\omega \ell }(r) q_{\omega \ell }(r)
- \frac {1}{rf^{\frac {1}{2}}} \right]
\right. \nonumber \\ & &  \left.
+ \frac {T}{4\pi } \sum _{\ell = 0}^{\infty } \left[ \left( 2 \ell + 1 \right)
C_{0 \ell } p_{0 \ell }(r) q_{0 \ell }(r)
- \frac {1}{rf^{\frac {1}{2}}} \right]
\right\} .
\nonumber \\
\label{eq:phi2unren}
\end{eqnarray}
To renormalize this expression, we subtract the usual divergent terms
$\langle \phi ^{2} \rangle _{\rm {div}}$ before taking the limit $\epsilon \rightarrow 0$
\cite{christensen}:
\begin{equation}
\langle \phi ^{2} \rangle _{\rm {div}}
  =  \frac {1}{8\pi ^{2} \sigma }
+ \frac {1}{8\pi ^{2}} \left[ m^{2} + \left( \xi - \frac {1}{6} \right) R \right]
\left[ C + \frac {1}{2}\ln \left( \frac {\mu ^{2} \left| \sigma \right| }{2} \right) \right]
- \frac {m^{2}}{16 \pi ^{2}}
+ \frac {1}{96\pi ^{2}}R_{\alpha \beta } \frac {\sigma ^{\alpha } \sigma ^{\beta }}{\sigma },
\label{eq:phi2div}
\end{equation}
where $\sigma ^{\alpha } = \sigma ^{;\alpha }$, the quantity $C$ is Euler's constant, and $R_{\alpha \beta }$ is the
(four-dimensional) Ricci tensor.
If we are considering a massive scalar field, then the constant $\mu $ is simply equal to $m$, the mass of the field.
However, for a massless scalar field, the constant $\mu $ is arbitrary \cite{ash}.
It corresponds in the massless case to a finite renormalization of terms in the gravitational action.

For our particular point-splitting, the quantities appearing in (\ref{eq:phi2div}) have already been computed
\cite{christensen,anderson,ash}:
\begin{eqnarray}
\sigma & = &
\frac {1}{2} f \epsilon ^{2} - \frac {1}{96} f \left( \frac {df}{dr} \right) ^{2} \epsilon ^{4} + O(\epsilon ^{6});
\nonumber \\
\sigma ^{\tau } & = &
- \epsilon +  \frac {1}{24} \left( \frac {df}{dr} \right) ^{2} \epsilon ^{3} + O(\epsilon ^{5});
\nonumber \\
\sigma ^{r} & = &
\frac {1}{4} f \frac {df}{dr} \epsilon ^{2} + O(\epsilon ^{4}) ;
\nonumber \\
\sigma ^{\theta } & = & \sigma ^{\varphi } = 0.
\end{eqnarray}
The subtraction terms (\ref{eq:phi2div}) then simplify to \cite{anderson}:
\begin{eqnarray}
\langle \phi ^{2} \rangle _{\rm {div}}
 & = &
 \frac {1}{4\pi ^{2} \epsilon ^{2} f}
% \nonumber \\ & &
+ \frac {1}{8\pi ^{2}} \left[ m^{2} + \left( \xi - \frac {1}{6} \right) R \right]
\left[ C + \frac {1}{2} \ln \left( \frac {\mu ^{2} f \epsilon ^{2}}{4} \right) \right]
%\nonumber \\ & &
- \frac {m^{2}}{16\pi ^{2}}
+\frac {1}{192\pi ^{2}f}\left( \frac {df}{dr} \right) ^{2}
\nonumber \\ & &
- \frac {1}{96\pi ^{2}} \frac {d^{2}f}{dr^{2}}
- \frac {1}{48\pi ^{2}r} \frac {df}{dr}.
\label{eq:phi2subtract}
\end{eqnarray}
In order to subtract (\ref{eq:phi2subtract}) from (\ref{eq:phi2unren}), we first need to write (\ref{eq:phi2subtract})
in terms of mode sums.
This is done using the following identities, valid for small $\epsilon $ and any $\kappa >0$ \cite{ash,howard}:
\begin{eqnarray}
\frac {1}{\epsilon ^{2}} & = &
-\kappa ^{2} \sum _{n=1}^{\infty } n \cos (n \kappa \epsilon ) - \frac {\kappa ^{2}}{12} + O(\epsilon ^{2});
\nonumber \\
-\frac{1}{2} \ln \left( \kappa ^{2} \epsilon ^{2} \right)
& = &
\sum _{n=1}^{\infty } \frac {\cos ( n \kappa \epsilon )}{n} + O(\epsilon ^{2}).
\end{eqnarray}
We choose $\kappa = 2\pi T$, so that $\omega $ in Eq. (\ref{eq:phi2unren}) is equal to $n\kappa $.
Then the subtraction terms (\ref{eq:phi2subtract}) become
\begin{eqnarray}
\langle \phi ^{2} \rangle _{\rm {div}}
 & = &
 -\frac {\kappa }{4\pi ^{2} f}  \sum _{n=1}^{\infty } \omega \cos (\omega \epsilon )
% \nonumber \\ & &
- \frac {\kappa }{8\pi ^{2}} \left[ m^{2} + \left( \xi - \frac {1}{6} \right) R \right]
 \sum _{n=1}^{\infty } \frac {\cos ( \omega \epsilon )}{\omega }
\nonumber \\ & &
+ \frac {1}{8\pi ^{2}} \left[ m^{2} + \left( \xi - \frac {1}{6} \right) R \right]
\left[ C + \frac {1}{2} \ln \left( \frac {\mu ^{2} f }{4\kappa ^{2}} \right) \right]
\nonumber \\ & &
- \frac {m^{2}}{16\pi ^{2}}
+\frac {1}{192\pi ^{2}f}\left( \frac {df}{dr} \right) ^{2}
- \frac {1}{96\pi ^{2}} \frac {d^{2}f}{dr^{2}}
%\nonumber \\ & &
- \frac {1}{48\pi ^{2}r} \frac {df}{dr} - \frac {\kappa ^{2}}{48 \pi ^{2}f} + O(\epsilon ^{2}).
\end{eqnarray}
This is now in a form suitable for subtracting from (\ref{eq:phi2unren}), and the limit $\epsilon \rightarrow 0$
can then be taken to give the final, renormalized, expectation value \cite{ash}:
\begin{equation}
\langle \phi ^{2} \rangle _{\rm {ren}} = \langle \phi ^{2} \rangle _{\rm {analytic}} +
\langle \phi ^{2} \rangle _{\rm {numeric}}
\end{equation}
where
\begin{eqnarray}
\langle \phi ^{2} \rangle _{\rm {analytic}} & = &
%\nonumber \\ & &
 \frac {m^{2}}{16\pi ^{2}}
-\frac {1}{192\pi ^{2}f}\left( \frac {df}{dr} \right) ^{2}
+ \frac {1}{96\pi ^{2}} \frac {d^{2}f}{dr^{2}}
%\nonumber \\ & &
+ \frac {1}{48\pi ^{2}r} \frac {df}{dr} + \frac {\kappa ^{2}}{48 \pi^{2}f}
\nonumber \\ & &
- \frac {1}{8\pi ^{2}} \left[ m^{2} + \left( \xi - \frac {1}{6} \right) R \right]
\left[ C + \frac{1}{2} \ln \left( \frac {\mu ^{2} f }{4\kappa ^{2}} \right) \right];
\label{eq:phi2analytic}
\\
\langle \phi ^{2} \rangle _{\rm {numeric}} & = &
 \frac {T}{2\pi } \sum _{n=1 }^{\infty } \left\{
\sum _{\ell = 0}^{\infty } \left[ \left( 2 \ell + 1 \right) C_{\omega \ell } p_{\omega \ell }(r) q_{\omega \ell }(r)
%\right. \right. \nonumber \\ & & \left. \left.
- \frac {1}{r{\sqrt {f}}} \right]
+ \frac {\omega }{f} +  \frac {1}{2\omega } \left[ m^{2} + \left( \xi - \frac {1}{6} \right) R \right]
\right\}
\nonumber \\ & &
+ \frac {T}{4\pi } \sum _{\ell = 0}^{\infty } \left[ \left( 2 \ell + 1 \right)
C_{0 \ell } p_{0 \ell }(r) q_{0 \ell }(r)
- \frac {1}{r{\sqrt {f}}} \right] .
\nonumber \\
\label{eq:phi2ren}
\end{eqnarray}
\end{widetext}
As the name suggests, $\langle \phi ^{2} \rangle _{\rm {analytic}}$ has a simple form which can be easily computed
for any $\kappa $ and metric function $f$.
On the other hand, the quantity $\langle \phi ^{2} \rangle _{\rm {numeric}}$ can, in general, only be computed
numerically as the mode equation (\ref{eq:modes}) needs to be integrated.
Solutions to the mode equation are not known in closed form for general $\omega $ even for Schwarzschild black holes.
One might hope that $\langle \phi ^{2} \rangle _{\rm {analytic}}$ is a good approximation to
$\langle \phi ^{2} \rangle _{\rm {ren}}$, at least in some region of the space-time.
This is discussed in \cite{ash} for the case of Reissner-Nordstr\"om black holes, and we will examine this issue
for lukewarm black holes in section \ref{sec:numerics}.

\subsection{Computation of $\langle \phi ^{2} \rangle _{\rm {numeric}}$}
\label{sec:WKB}

Before we can compute $\langle \phi ^{2} \rangle _{\rm {numeric}}$, further work is needed.
The mode sums in $\langle \phi ^{2} \rangle _{\rm {numeric}}$ converge so slowly as to render their numerical
computation impractical.
We therefore employ a WKB-like approximation, which will give the large $\omega $, $\ell $ behaviour of the mode
sums in (\ref{eq:phi2ren}).
We have found it simplest to use the WKB approach of Howard \cite{howard}, which is different from that
used in Ref. \cite{anderson,ash}.

Define a new function $\beta _{\omega \ell }(r)$ for $\omega \neq 0$ by
\begin{equation}
\beta _{\omega \ell }(r) = C_{\omega \ell } p_{\omega \ell }(r) q_{\omega \ell }(r).
\end{equation}
Then $\beta _{\omega \ell }$ satisfies the following differential equation (cf. \cite{howard}):
\begin{equation}
\beta _{\omega \ell}  =
\frac {1}{2\chi _{\omega \ell }} \left[
1 - \frac {1}{\chi _{\omega \ell }^{2}} \left(
\frac {1}{{\sqrt {\beta _{\omega \ell }}}} \frac {d^{2}({\sqrt {\beta _{\omega \ell }}})}{d\zeta ^{2}}
- \eta  \right) \right] ^{-\frac {1}{2}},
\label{eq:betaDE}
\end{equation}
where we have defined a new independent variable $\zeta $ by
\begin{equation}
\frac {d}{d\zeta } = r^{2} f \frac {d}{dr},
\end{equation}
and the functions $\chi _{\omega \ell }(r)$ and $\eta (r)$ are given by
\begin{eqnarray}
\chi _{\omega \ell }(r) & = &
{\sqrt { \omega ^{2} r^{4} + \left( \ell + \frac {1}{2} \right) ^{2} r^{2} f }}
\nonumber \\
\eta (r) & = &
- \frac {1}{4} fr^{2} + \left( m^{2} + \xi R \right) fr^{4}.
\label{eq:chi}
\end{eqnarray}
We are looking for the behaviour of $\beta _{\omega \ell }(r)$ when either $\omega $ or $\ell $ are large, that is,
when $\chi _{\omega \ell }(r)$ is large.
This is found by inserting a fictitious parameter $\varepsilon $ into (\ref{eq:betaDE}) as follows:
\begin{equation}
\beta _{\omega \ell}  =
\frac {1}{2\chi _{\omega \ell }} \left[
1 - \frac {1}{\varepsilon ^{2} \chi _{\omega \ell }^{2}} \left(
\frac {1}{{\sqrt {\beta _{\omega \ell }}}} \frac {d^{2}({\sqrt {\beta _{\omega \ell }}})}{d\zeta ^{2}}
- \eta  \right) \right] ,
\label{eq:betaDEepsilon}
\end{equation}
and then expanding $\beta _{\omega \ell }$ in inverse powers of $\varepsilon $:
\begin{equation}
\beta _{\omega \ell }(r) = \beta _{0 \omega \ell }(r) + \varepsilon ^{-2} \beta _{1 \omega \ell }(r)
+ \varepsilon ^{-4} \beta _{2\omega \ell }(r) + \ldots ,
\end{equation}
finally setting $\varepsilon =1$ at the end of the calculation.

It is straightforward to read off from (\ref{eq:betaDEepsilon}) that $\beta _{0\omega \ell }(r)$ has the simple
form
\begin{equation}
\beta _{0\omega \ell }(r) = \frac {1}{2\chi _{\omega \ell }(r)}.
\label{eq:beta0}
\end{equation}
The next term in the expansion $\beta _{1\omega \ell }(r)$ is more complicated but not difficult to calculate:
\begin{equation}
\beta _{1\omega \ell }(r) =
- \frac {r^{6}f}{64 \chi _{\omega \ell }^{7}}
\left( A_{1} \chi _{\omega \ell }^{4} + B_{1} \chi _{\omega \ell }^{2} + C_{1} \right) ,
\label{eq:beta1}
\end{equation}
where
\begin{eqnarray}
A_{1} & = &
\frac {1}{f^{2} r^{4}}
\left[
-f {r}^{2} \left( {\frac {df}{dr}}  \right) ^{2}
+12 f ^{2} r \frac {df}{dr}
+4 f ^{3}
-4f ^{2}
\right. \nonumber \\ & & \left.
+16{r}^{2}{\xi}R f ^{2}
+4f ^{2}{r}^{2}{\frac {d^{2}f}{d{r}^{2}}}
+16{r}^{2}{m}^{2}f^{2}
\right] ;
\nonumber \\
B_{1} & = &
\frac {\omega^{2}}{f^{2}}
\left[ -4 f^{2} r^{2} {\frac {d^{2}f}{d{r}^{2}}}
+6f {r}^{2} \left( {\frac {df}{dr}} \right) ^{2}
\right. \nonumber \\ & & \left.
 -16  f^{2}r {\frac {df}{dr}}
 +16 f ^{3}
 \right] ;
\nonumber \\
C_{1} & = &
-5 \frac {\omega ^{4}{r}^{4}}{f}
\left[ 4 f^{2}+{r}^{2} \left( {\frac {df}{dr}}\right) ^{2}
-4fr{\frac {df}{dr}}  \right] .
\end{eqnarray}
It can be seen from (\ref{eq:beta0},\ref{eq:beta1}) that, for large $\chi _{\omega \ell }$ (or, equivalently, for
large $\omega $ or $\ell $ if $f\neq 0$),
\begin{equation}
\beta _{0\omega \ell } \sim \chi _{\omega \ell }^{-1}, \qquad
\beta _{1\omega \ell }\sim \chi _{\omega \ell }^{-3},
\end{equation}
and we find similarly that $\beta _{2\omega \ell } \sim \chi _{\omega \ell }^{-5}$ etc.
If we consider the sum
\begin{equation}
 \sum _{n=1 }^{\infty }
\sum _{\ell = 0}^{\infty } \left( 2 \ell + 1 \right) \left[  C_{\omega \ell } p_{\omega \ell }(r) q_{\omega \ell }(r)
- \beta _{0\omega \ell }(r) - \beta _{1\omega \ell }(r) \right] ,
\end{equation}
then the summand is $O(\ell ^{-4})$ for large $\ell $.
This is the behaviour we observe when we compute the summand numerically in section \ref{sec:mode}.
When this is summed over $\ell $, we obtain a summand which is $O(\omega ^{-3})$ and therefore converges rapidly.
Therefore it is sufficient to subtract just $\beta _{0\omega \ell }$ and $\beta _{1\omega \ell }$ from the mode
sums in $\langle \phi ^{2} \rangle _{\rm {numeric}}$ (\ref{eq:phi2ren}).

The contribution to (\ref{eq:phi2ren}) from the $\omega = 0$ modes needs to be considered separately.
When $\omega =0$, from (\ref{eq:chi}) we have $\chi _{0\ell }= \left( \ell + \frac {1}{2} \right) rf^{\frac {1}{2}}$
and therefore (\ref{eq:beta0},\ref{eq:beta1}) become
\begin{eqnarray}
\beta _{0 0 \ell } (r) & = & \frac {1}{2rf^{\frac {1}{2}}}\left( \ell + \frac {1}{2} \right) ^{-1} ;
\nonumber \\
\beta _{1 0 \ell } (r) & = & - \frac {r^{3}}{64f^{\frac {1}{2}}} A_{1}
\left( \ell + \frac {1}{2} \right) ^{-3} .
\label{eq:betaoneomega0}
\end{eqnarray}
Therefore $(2\ell + 1)\beta _{0 0 \ell }(r) = \left(rf^{\frac {1}{2}}\right)^{-1}$ and this part of the WKB
approximation has already been subtracted from the sum.
In this case we consider
\begin{equation}
\sum _{\ell = 0}^{\infty } \left( 2 \ell + 1 \right) \left[  C_{0 \ell } p_{0 \ell }(r) q_{0 \ell }(r)
- \beta _{00 \ell }(r) - \beta _{10 \ell }(r) \right] ,
\end{equation}
and again the summand is $O(\ell ^{-4})$ for large $\ell $ as required.

\begin{widetext}
We therefore write
\begin{eqnarray}
\langle \phi ^{2} \rangle _{\rm {numeric}} & = &
 \frac {T}{2\pi } \sum _{n=1 }^{\infty } \left\{
\sum _{\ell = 0}^{\infty } \left( 2 \ell + 1 \right) \left[ C_{\omega \ell } p_{\omega \ell }(r) q_{\omega \ell }(r)
%\right. \right. \nonumber \\ & & \left. \left.
- \beta _{0\omega \ell }(r) - \beta _{1\omega \ell }(r) \right]
\right. \nonumber \\ & & \left.
+ \sum _{\ell = 0}^{\infty } \left[  \left( 2\ell + 1 \right)
\left[ \beta _{0\omega \ell } (r) + \beta _{1\omega \ell }(r) \right]
- \frac {1}{rf^{\frac {1}{2}}} \right]
%\right. \nonumber \\ & & \left.
+ \frac {\omega }{f} +  \frac {1}{2\omega } \left[ m^{2} + \left( \xi - \frac {1}{6} \right) R \right]
\right\}
\nonumber \\ & &
+ \frac {T}{4\pi } \left\{ \sum _{\ell = 0}^{\infty } \left( 2 \ell + 1 \right)
\left[ C_{0 \ell } p_{0 \ell }(r) q_{0 \ell }(r)
- \beta _{00 \ell}(r) - \beta _{10 \ell }(r)  \right]
+ \sum _{\ell =0}^{\infty } \left( 2\ell + 1 \right) \beta _{10\ell }(r) \right\},
%\nonumber \\
\label{eq:phi2numbeta}
\end{eqnarray}
where we have used the fact that $(2\ell + 1)\beta _{0 0 \ell }(r) = \left(rf^{\frac {1}{2}}\right)^{-1}$.
The first and last lines in (\ref{eq:phi2numbeta}) are now amenable to numerical computation.
The final sum is readily computed using (\ref{eq:betaoneomega0}) to be
\begin{equation}
\sum _{\ell =0}^{\infty } \left( 2\ell + 1 \right) \beta _{10\ell }(r)
= - \frac {r^{3}}{64f^{\frac {1}{2}}} A_{1}
\sum _{\ell = 0}^{\infty } 2\left( \ell + \frac {1}{2} \right) ^{-2}
=- \frac {\pi ^{2}r^{3}}{64f^{\frac {1}{2}}} A_{1} .
\end{equation}

Since we have analytic expressions for $\beta _{0\omega \ell }(r)$ and $\beta _{1\omega \ell }(r)$,
we next examine the term
\begin{equation}
 \sum _{n=1 }^{\infty } \left\{ \sum _{\ell = 0}^{\infty }
\left( 2\ell + 1 \right)
\left[ \left( \beta _{0\omega \ell (r)} - \frac {1}{rf^{\frac {1}{2}}} \right)  + \beta _{1\omega \ell }(r)
 \right]
+ \frac {\omega }{f} +  \frac {1}{2\omega } \left[ m^{2} + \left( \xi - \frac {1}{6} \right) R \right]
\right\} .
\end{equation}
The sums over $\ell $ are most easily found using the Watson-Sommerfeld formula (see, for example, \cite{jo}),
valid for any function analytic in the right-hand half plane:
\begin{equation}
\sum _{\ell =0}^{\infty } {\cal {F}}(\ell )
 =  \int _{0}^{\infty } {\cal {F}} \left( \lambda - \frac {1}{2} \right) \, d\lambda
- \Re \left[ i \int _{0}^{\infty } \frac {2}{1+e^{2\pi \lambda } } {\cal {F}} \left( i\lambda - \frac {1}{2} \right)
\, d\lambda \right] .
\label{eq:watson}
\end{equation}
Using (\ref{eq:watson}), we write, for $\omega \neq 0$,
\begin{eqnarray}
\sum _{\ell = 0}^{\infty }
\left( 2\ell + 1 \right) \left[ \beta _{0\omega \ell (r)} - \frac {1}{rf^{\frac {1}{2}}} \right]
& = & I_{0}(\omega ,r) + {\tilde {J}}_{0}(\omega ,r) ;
\nonumber \\
\sum _{\ell = 0}^{\infty } \left( 2\ell + 1 \right)  \beta _{1\omega \ell (r)}
& = & I_{1}(\omega ,r) + J_{1}(\omega ,r) ;
\end{eqnarray}
where the $I_{i}(\omega , r)$ are the integrals from the first term in (\ref{eq:watson}) and the
$J_{i}(\omega, r)$ are the integrals from the second term in (\ref{eq:watson}).
The $I_{i}(\omega ,r)$ integrals are easily calculated for $\omega \neq 0$ (here $\ell = \lambda - 1/2$):
\begin{eqnarray}
I_{0}(\omega ,r ) & = &
\int _{0}^{\infty } \left[ 2\lambda \beta _{0\omega \ell }(r) - \frac {1}{rf^{\frac {1}{2}}} \right] \, d\lambda
= - \frac {\omega }{f} ;
 \nonumber \\
 I_{1}(\omega ,r) & = &
 \int _{0}^{\infty } 2\lambda \beta _{1\omega \ell }(r) \, d\lambda
=
  -\frac {1}{2\omega } \left[ m^{2} + \left( \xi - \frac {1}{6}\right) R \right] - \frac {1}{24r^{2}\omega }.
\end{eqnarray}
Substituting these results in (\ref{eq:phi2numbeta}) gives
\begin{eqnarray}
\langle \phi ^{2} \rangle _{\rm {numeric}} & = &
 \frac {T}{2\pi } \sum _{n=1 }^{\infty } \left\{
\sum _{\ell = 0}^{\infty } \left( 2 \ell + 1 \right) \left[ C_{\omega \ell } p_{\omega \ell }(r) q_{\omega \ell }(r)
%\right. \right. \nonumber \\ & & \left. \left.
- \beta _{0\omega \ell }(r) - \beta _{1\omega \ell }(r) \right]
%\right. \nonumber \\ & & \left.
+ {\tilde {J}}_{0}(\omega, r) + J_{1}(\omega ,r ) - \frac {1}{24r^{2} \omega }
\right\}
\nonumber \\ & &
+ \frac {T}{4\pi } \left\{ \sum _{\ell = 0}^{\infty } \left( 2 \ell + 1 \right)
\left[ C_{0 \ell } p_{0 \ell }(r) q_{0 \ell }(r) - \beta _{00\ell }(r) - \beta _{10\ell }(r)
 \right]
 - \frac {\pi ^{2}r^{3}}{64f^{\frac {1}{2}}} A_{1} \right\}.
%\nonumber \\
\label{eq:phi2numbetaone}
\end{eqnarray}

Next we examine the $J_{i}(\omega , r)$ integrals (here $\ell = i\lambda - 1/2$):
\begin{equation}
{\tilde {J}}_{0}(\omega ,r) = - \Re \left\{ i \int _{0}^{\infty } \frac {2}{1+e^{2\pi \lambda }}
\left[ 2\lambda i \beta _{0\omega \ell }(r) - \frac {1}{rf^{\frac {1}{2}}} \right] \, d\lambda \right\},
 \qquad
 J_{1}(\omega ,r ) =  \Re \left\{  \int _{0}^{\infty } \frac {4\lambda }{1+e^{2\pi \lambda }}
 \beta _{1\omega \ell }(r)  \, d\lambda \right\} .
\end{equation}
We consider each of these in turn.
Firstly, for ${\tilde {J}}_{0}(\omega , r)$, substituting in the form of
$\beta _{0\omega \ell }(r)$ (\ref{eq:beta0}) gives
\begin{equation}
{\tilde {J}}_{0}(\omega , r) =  \frac {1}{rf^{\frac {1}{2}}} \Re \left[
\int _{0}^{\infty } \frac {2i}{1+e^{2\pi \lambda }} \, d\lambda
+ \int _{0}^{a} \frac {2\lambda }{\left( 1 + e^{2\pi \lambda }\right) \left( a^{2}-\lambda ^{2} \right) ^{\frac {1}{2}}}
\, d\lambda
+ \int _{a}^{\infty }
\frac {2\lambda }{\left( 1 + e^{2\pi \lambda }\right) \left( a^{2}-\lambda ^{2} \right)  ^{\frac {1}{2}}}
\, d\lambda \right] ,
\label{eq:J0one}
\end{equation}
where we have defined
\begin{equation}
a = \frac {\omega r}{f^{\frac {1}{2}}}.
\label{eq:adef}
\end{equation}
The first and third integrals in (\ref{eq:J0one}) do not contribute, leaving just the second integral.
The integrand is integrable but not regular at $\lambda = a$.
Furthermore, this integral is not known in closed form and will need to be computed numerically.
It is easier numerically to have a regular integrand, so we integrate once by parts to obtain
\begin{equation}
{\tilde {J}}_{0}(\omega , r) = \frac {\omega }{f} - \frac {4\pi \omega }{f}
\int _{0}^{a} \left( 1 - \frac {\lambda ^{2}}{a^{2}} \right) ^{\frac {1}{2}}
\frac {e^{2\pi \lambda }}{\left( 1 + e^{2\pi \lambda }\right) ^{2} } \, d\lambda.
\end{equation}
For large $\omega $, this integral behaves like
\begin{equation}
{\tilde {J}}_{0}(\omega ,r ) = \frac {1}{24\omega r^{2}} + O(\omega ^{-3}) .
\end{equation}
We therefore take the $-(24\omega r^{2})^{-1}$ term from (\ref{eq:phi2numbetaone}) and consider instead
\begin{equation}
J_{0}(\omega , r) = \frac {\omega }{f} - \frac {1}{24r^{2}\omega }
- \frac {4\pi \omega }{f}
\int _{0}^{a} \left( 1 - \frac {\lambda ^{2}}{a^{2}} \right) ^{\frac {1}{2}}
\frac {e^{2\pi \lambda }}{\left( 1 + e^{2\pi \lambda }\right) ^{2} } \, d\lambda
 \sim O(\omega ^{-3})
\label{eq:J0}
\end{equation}
as $\omega \rightarrow \infty $.
This means that the sum $\sum _{n=1}^{\infty } J_{0}(\omega ,r)$ can be computed separately from
the other terms in $\langle \phi ^{2} \rangle _{\rm {numeric}}$ (\ref{eq:phi2numbetaone}).

The integral $J_{1}(\omega ,r)$ is more complicated.
Using the form (\ref{eq:beta1}) of $\beta _{1\omega \ell }(r)$, we find:
\begin{equation}
J_{1}(\omega ,r ) = -\frac {1}{16rf^{\frac {1}{2}}} \left[
r^{4} A_{1} K_{1}(\omega ,r) + \frac {1}{\omega ^{2}} B_{2} K_{2}(\omega ,r)
+ \frac {1}{\omega ^{4}r^{4}} C_{1} K_{3} (\omega ,r) ,
\right] ,
\label{eq:J1}
\end{equation}
where we have defined new integrals
\begin{eqnarray}
K_{1}(\omega ,r) & = &
\frac {1}{a} \Re \left[  \int _{0}^{\infty } \frac {q}{\left( 1 - q^{2} \right) ^{\frac {3}{2}}
\left( 1 + e^{2\pi a q } \right) } \, dq \right] ;
\nonumber \\
K_{2} (\omega ,r ) & = &
\frac {1}{a} \Re \left[  \int _{0}^{\infty } \frac {q}{\left( 1 - q^{2} \right) ^{\frac {5}{2}}
\left( 1 + e^{2\pi a q } \right) } \, dq \right] ;
\nonumber \\
K_{3}(\omega ,r) & = &
\frac {1}{a} \Re \left[  \int _{0}^{\infty } \frac {q}{\left( 1 - q^{2} \right) ^{\frac {7}{2}}
\left( 1 + e^{2\pi a q } \right) } \, dq \right] ;
\label{eq:Kint}
\end{eqnarray}
and $q=\lambda /a$.
The integrands in all three integrals have branch points at $q=\pm 1$, so we cut the plane along the interval
$[-1,1]$ and consider the contour shown in Fig. \ref{fig:contour}.
\begin{figure}[h]
\begin{center}
\includegraphics[width=4cm,angle=270]{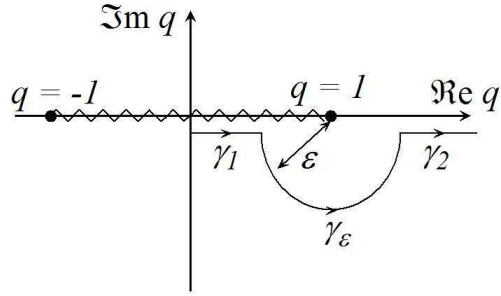}
\end{center}
\caption{Contour used for computing the integrals $K_{i}(\omega ,r)$ (\ref{eq:Kint}).}
\label{fig:contour}
\end{figure}
\newline
The contribution to each $K_{i}(\omega ,r)$ from the contour $\gamma _{2}$ is zero, so we write, for each $i=1,2,3$:
\begin{equation}
K_{i}(\omega ,r) = \frac {1}{a} \lim _{\varepsilon \rightarrow 0} \left[ L_{i}(\omega ,r) + M_{i}(\omega ,r) \right]
\end{equation}
where each $L_{i}(\omega ,r )$ is the contribution from the contour $\gamma _{1}$ and $M_{i}(\omega ,r )$ is the
contribution from the contour $\gamma _{\varepsilon }$ in Fig. \ref{fig:contour}.
We illustrate the procedure for calculating these by considering $L_{1}(\omega ,r)$ and $M_{1}(\omega ,r )$.
The method works similarly for $K_{2}(\omega ,r )$ and $K_{3}(\omega ,r)$ but is more complicated.

Considering $L_{1}(\omega ,r )$ first, we have
\begin{equation}
L_{1}(\omega ,r ) = \Re \left[ \int _{\gamma _{1}} \frac {q}{\left( 1 - q^{2} \right) ^{\frac {3}{2}}
\left( 1 + e^{2\pi a q } \right) } \, dq \right]
= \int _{0}^{1-\varepsilon } \frac {q}{\left( 1 - q^{2} \right) ^{\frac {3}{2}}
\left( 1 + e^{2\pi a q } \right) } \, dq .
\end{equation}
Integrating by parts gives
\begin{equation}
L_{1} (\omega ,r ) = \frac {1}{2{\sqrt {\varepsilon }} \left( 1 + e^{2\pi a} \right) }
- \frac {1}{2} + 2\pi a \int _{0}^{1-\varepsilon } \frac {\left( 1 - q^{2} \right) ^{-\frac {1}{2}}
e^{2\pi aq} }{\left( 1 + e^{2\pi a q} \right) ^{2}} \, dq + O(\varepsilon ^{\frac {1}{2}}).
\label{eq:L1int}
\end{equation}
Note that the first term in (\ref{eq:L1int}) is divergent as $\varepsilon \rightarrow 0$, and that we
have now isolated a finite integral (although the integrand is not regular).
We next turn to $M_{1}(\omega ,r )$:
\begin{equation}
M_{1}(\omega ,r ) = \Re \left[ \int _{\gamma _{\varepsilon }} \frac {q}{\left( 1 - q^{2} \right) ^{\frac {3}{2}}
\left( 1 + e^{2\pi a q } \right) } \, dq \right]  .
\end{equation}
Along $\gamma _{\varepsilon }$, we have
\begin{equation}
q -1 = -\varepsilon e^{i\vartheta }, \qquad 0 < \vartheta < \pi .
\end{equation}
Changing the variable of integration in $M_{1}(\omega ,r)$ to $\vartheta $, and computing the integral, we find
\begin{equation}
M_{1}(\omega ,r ) = -\frac {1}{2{\sqrt {\varepsilon }} \left( 1 + e^{2\pi a} \right) } + O(\varepsilon ^{\frac {1}{2}}).
\end{equation}
Therefore, adding $L_{1}(\omega ,r)$ and $M_{1}(\omega ,r )$ will give a finite quantity and we can take the limit
$\varepsilon \rightarrow 0$.
It is helpful for numerical computation to integrate by parts again to give a regular integrand:
\begin{equation}
K_{1}(\omega ,r) = - \frac {1}{2a} + \frac {\pi ^{2}e^{2\pi a}}{\left( 1 + e^{2\pi a} \right) ^{2}}
+ 4\pi ^{2} \int _{0}^{a} \sin ^{-1} \left( \frac {\lambda }{a} \right) \frac {e^{2\pi \lambda }\left(
e^{2\pi \lambda } -1 \right)  }{\left( 1 + e^{2\pi \lambda } \right) ^{3} }\, d\lambda ,
\label{eq:K1}
\end{equation}
where we have returned to the original variable $\lambda = aq$.

The calculation proceeds similarly for $K_{2}(\omega ,r)$ and $K_{3}(\omega ,r)$, except that more integrations by parts
are required. In each case, the divergences in $L_{i}(\omega ,r)$ and $M_{i}(\omega ,r )$ cancel to give the following
finite quantities:
\begin{eqnarray}
K_{2}(\omega ,r ) & =&
-\frac {1}{6a} + \frac {8\pi ^{3}a}{3} \int _{0}^{a} \left( 1 - \frac {\lambda ^{2}}{a^{2}}  \right) ^{\frac {1}{2}}
\frac {e^{2\pi \lambda } \left( -1 - e^{4\pi \lambda } + 4e^{2\pi \lambda } \right)}{\left(
1 + e^{2\pi \lambda } \right) ^{4} } \, d\lambda ;
\label{eq:K2}
\\
K_{3}(\omega ,r ) & = &
-\frac {1}{10a}
- \frac {16\pi ^{4}a}{15} \int _{0}^{a} \lambda  \left( 1 - \frac {\lambda ^{2}}{a^{2}} \right) ^{\frac {1}{2}}
\frac {e^{2\pi \lambda }\left( -1 + 11 e^{2\pi \lambda } - 11 e^{4\pi \lambda } + e^{6\pi \lambda } \right)}{
\left( 1 + e^{2\pi \lambda } \right) ^{5}} \, d\lambda .
\label{eq:K3}
\end{eqnarray}
It is straightforward to verify that, for each $i$, the integrals $K_{i}(\omega ,r)$ are $O(\omega ^{-3})$ for
large $\omega $, so that the sums over $n$ required in (\ref{eq:phi2numbetaone}) will converge rapidly.

At this stage it is helpful to bring together our results for $\langle \phi ^{2} \rangle _{\rm {numeric}}$.
From (\ref{eq:phi2numbetaone}) we have
\begin{equation}
\langle \phi ^{2} \rangle _{\rm {numeric}}  =
\Sigma
+ \frac {T}{2\pi } \sum _{n=1}^{\infty } \left[  J_{0}(\omega, r) + J_{1}(\omega ,r ) \right],
\label{eq:phi2numsplit}
\end{equation}
where
\begin{eqnarray}
\Sigma & = &  \frac {T}{2\pi } \sum _{n=1 }^{\infty } \left\{
\sum _{\ell = 0}^{\infty } \left( 2 \ell + 1 \right) \left[ C_{\omega \ell } p_{\omega \ell }(r) q_{\omega \ell }(r)
- \beta _{0\omega \ell }(r) - \beta _{1\omega \ell }(r) \right] \right\}
\nonumber \\ & &
+ \frac {T}{4\pi } \left\{ \sum _{\ell = 0}^{\infty }  \left( 2 \ell + 1 \right) \left[
C_{0 \ell } p_{0 \ell }(r) q_{0 \ell }(r)
- \beta _{00\ell}(r) - \beta _{10\ell }(r) \right] - \frac {\pi ^{2}r^{3}}{64f^{\frac {1}{2}}} A_{1} \right\} ,
\label{eq:modesum}
\end{eqnarray}
and the numerical integrals $J_{0}(\omega ,r)$ and $J_{1}(\omega ,r)$ are given by (\ref{eq:J0}) and
(\ref{eq:J1}) respectively (with the $K_{i}(\omega ,r)$ given by (\ref{eq:K1}--\ref{eq:K3})).
In the following we shall refer to $\Sigma $ as the `mode sum'.
%\end{widetext}

\section{Numerical Results for Lukewarm Black Holes}
\label{sec:numerics}

We now turn to the computation of $\langle \phi ^{2} \rangle _{\rm {ren}}$ for a specific example, namely
lukewarm black holes \cite{romans}.
We briefly review the key features of these black holes before describing the results of our calculation.
For many of our numerical computations we have used standard routines \cite{NR}.
Therefore we omit much of the detail of the numerical methods used, which are comprehensively discussed in
Ref. \cite{PMYthesis}.

\end{widetext}

\subsection{Lukewarm Black Holes}
\label{sec:lukewarm}

Lukewarm black holes are a particular type of Reissner-Nordstr\"om-de Sitter space-time, with metric given
by (\ref{eq:metric},\ref{eq:fRNdS}) with $M=Q$.
For $4M<{\sqrt {3/\Lambda }}$, there are three distinct horizons (a black hole event horizon at $r=r_{+}$, an inner
(Cauchy) horizon at $r=r_{-}$ and a cosmological horizon at $r=r_{c}$), given by
\begin{eqnarray}
r_{-} & = & \frac {L}{2} \left[ -1 + {\sqrt {1+ \frac {4M}{L}}} \right] ;
\nonumber \\
r_{+} & = &  \frac {L}{2} \left[ 1 - {\sqrt {1 - \frac {4M}{L}}} \right] ;
\nonumber \\
r_{c} & = & \frac {L}{2} \left[ 1 + {\sqrt {1 - \frac {4M}{L}}} \right] ;
\end{eqnarray}
where
\begin{equation}
L = {\sqrt {\frac {3}{\Lambda }}}, \qquad
4M < L .
\end{equation}
The Penrose diagram for this space-time can be found in \cite{mellor}.

In this case the event and cosmological horizons have equal surface gravities:
\begin{equation}
\kappa _{+} = \kappa _{c} = \frac {1}{L} {\sqrt {1 - \frac {4M}{L}}},
\end{equation}
which means that they have the same temperature $T=\kappa _{+}/2\pi $.
Here we are interested in the region between the event and cosmological horizons, which has a regular Euclidean section,
with topology $S^{2}\times S^{2}$ \cite{mellor}.
Our computations are for a thermal state at the natural temperature $T$.

In the following sections, we show plots of the constituent parts of $\langle \phi ^{2} \rangle _{\rm {ren}}$
for the specific case of $M=Q=0.1L$, although the main features are the same for other values of $M$.
For the remainder of this section, all dimensionful quantities ($r$, $M$, etc) are given in units of $L$.

\subsection{Mode Sum}
\label{sec:mode}

We begin our numerical analysis by calculating the mode sum (\ref{eq:modesum}).
Firstly we need to find the modes themselves, by integrating the mode equation (\ref{eq:modes}) which
is satisfied by $p_{\omega \ell }(r)$ and $q_{\omega \ell }(r)$.
This is done using standard shooting techniques \cite{NR}.

\begin{figure}[h]
\begin{center}
\includegraphics[width=6cm,angle=270]{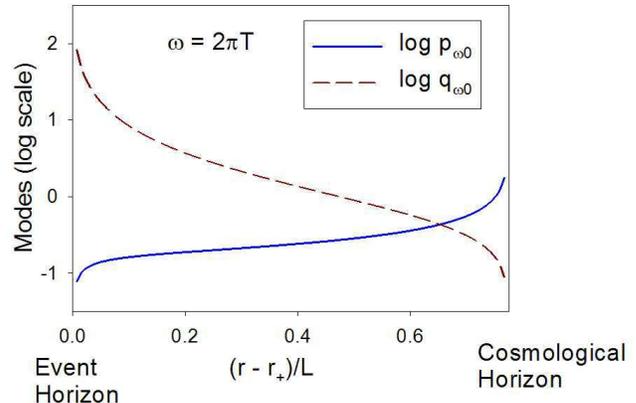}
\end{center}
\caption{Examples of mode solutions of the equation (\ref{eq:modes}), with $\omega = 2\pi T$ and $\ell = 0$.
The function $p_{\omega \ell }(r)$ vanishes at the event horizon (at the left-hand-edge of the plot)
and diverges at the cosmological horizon (at the right-hand-edge of the plot),
while the mode function $q_{\omega \ell }(r)$ vanishes at the cosmological horizon and diverges at the
event horizon.
For these mode functions, we have taken the scalar field to be massless and conformally coupled,
and the black hole metric (\ref{eq:metric},\ref{eq:fRNdS}) has parameter $M=Q=0.1L$.
The radial co-ordinate on the horizontal axis is in units of $L$.}
\label{fig:modes}
\end{figure}

The mode equation (\ref{eq:modes}) has regular singular points at the event and cosmological horizons.
Using the standard Frobenius method, we write the mode functions as power series:
\begin{equation}
S_{\omega \ell } = \sum _{i=0}^{\infty } a_{i} x^{i+\nu }
\end{equation}
where $S_{\omega \ell }$ is either $p_{\omega \ell }(r)$ or $q_{\omega \ell }(r)$, and $x=r-r_{+}>0$ if we
are considering the behaviour near the event horizon ($x=r_{c}-r>0$ if we are considering the
behaviour near the cosmological horizon).
By looking at the lowest order term in (\ref{eq:modes}) we obtain the indicial equation \cite{anderson}:
\begin{equation}
\nu ^{2} = \omega ^{2}\left( \left. \frac {df}{dr}\right| _{r=r_{0}} \right) ^{-2},
\end{equation}
where $r_{0}$ is either $r_{+}$ or $r_{c}$ as applicable.
In our mode sum (\ref{eq:modesum}), we are only summing over modes for which
\begin{equation}
\omega = 2\pi n T= n\kappa _{+}= \frac {n}{2}\left. \frac {df}{dx}\right| _{r=r_{0}}, \qquad n = 0,1,2,\ldots
\end{equation}
Therefore $\nu =\pm n/2$, and the roots of the indicial equation differ by an integer ($n>0$) or zero ($n=0$).
Therefore the linearly independent solutions of the mode equation (\ref{eq:modes}) are:
\begin{eqnarray}
S_{1\omega \ell } & = & \sum _{i=0}^{\infty } a_{i} x^{i+\frac {n}{2} } ;
\nonumber \\
S_{2\omega \ell } & = & \sum _{i=0}^{\infty } b_{i} x^{i-\frac {n}{2}}
+ {\cal {K}}_{\omega \ell } S_{1\omega \ell } \ln \left( \frac {x}{r_{0}} \right) ;
\label{eq:Frobeniusmodes}
\end{eqnarray}
where ${\cal {K}}_{\omega \ell }$ is a constant which definitely does not vanish if $n=0$ but may possibly vanish
if $n>0$, and $r_{0}=r_{+}$ or $r_{c}$ as applicable.
The mode functions are chosen so that $p_{\omega \ell }$ is regular (that is, has the form $S_{1\omega \ell }$)
near the event horizon, while $q_{\omega \ell }$ is regular near the cosmological horizon.
From (\ref{eq:Frobeniusmodes}), it is clear that $q_{\omega \ell }$ will diverge (that is, have the form
$S_{2\omega \ell }$) near the event horizon, and $p_{\omega \ell }$ will diverge near the cosmological horizon.
This can be seen in the examples of mode functions plotted in Fig. \ref{fig:modes}.

For $n>0$, it is clear that the mode functions $p_{\omega \ell }(r)$ will vanish at the event horizon and
that $q_{\omega \ell }(r)$ will vanish at the cosmological horizon.
However, for $n=0$, we have that $p_{0\ell }(r)$ is regular but non-zero at the event horizon (and similarly
for $q_{0\ell }(r)$ at the cosmological horizon).
In this case $p_{0\ell }(r)$ will still diverge at the cosmological horizon because of the $\ln (x/r_{0})$ term
in $S_{2\omega \ell }$ (\ref{eq:Frobeniusmodes}) (note that ${\cal {K}}_{0\ell }$ cannot be zero because there
must be two linearly independent solutions of the mode equation).
It is straightforward to show, from the mode equation (\ref{eq:modes}), that the mode functions are monotonic and
do not have zeros except possibly at a horizon.

To find the $p_{\omega \ell }(r)$ mode functions,
we start the numerical integration of the mode equation (\ref{eq:modes}) just
outside the event horizon, using as many terms in the power series expansion (\ref{eq:Frobeniusmodes}) of
$S_{1\omega \ell }$ as required for the desired accuracy.
As in \cite{anderson}, there is a complicated recurrence relation (involving seven terms in general) for
the coefficients in the power series expansion (\ref{eq:Frobeniusmodes}), which we do not reproduce here.
Starting with $a_{0}=1$, this recurrence relation is used to compute the power series expansion.
We then integrate outwards towards the cosmological horizon.
For the $q_{\omega \ell }(r)$ mode functions, we start integrating just inside the cosmological horizon, and
integrate towards the event horizon.
We use the Bulirsch-Stoer method \cite{NR} of integrating the differential equation (\ref{eq:modes}) because
the mode functions are not oscillating and because of the high degree of accuracy required.

Looking at the mode sum (\ref{eq:modesum}), it can be seen that for large $\omega $ or $\ell $, the rapid
convergence of the sum is dependent on subtracting very nearly equal quantities.
Therefore it is imperative to calculate the mode functions $p_{\omega \ell }(r)$,  $q_{\omega \ell }(r)$
with great accuracy.
We therefore used quadruple precision throughout our calculations.
The normalization constant $C_{\omega \ell }$ is computed from (\ref{eq:Wronskian}).
The constancy of the $C_{\omega \ell }$ as calculated from (\ref{eq:Wronskian}) for different values of $r$
represents a good check on the accuracy of our results.
For the particular case of $Q=M=0.1L$, we find that $C_{\omega \ell }$ remains constant to within
$10^{-26}\left| C_{\omega \ell }\right| $ for $0\le n \le 25$ and $0\le \ell \le 600$, which gives us enough
modes to get good convergence of the mode sum (\ref{eq:modesum}).

Once we have the mode functions $p_{\omega \ell }(r)$,  $q_{\omega \ell }(r)$, as the WKB approximants
$\beta _{0\omega \ell }(r)$ (\ref{eq:beta0}) and $\beta _{1\omega \ell }(r)$ (\ref{eq:beta1})
are known analytically, we are able to find the mode sum (\ref{eq:modesum}).
We found that the summand was $O(\ell ^{-4})$ for large $\ell $ as predicted in section \ref{sec:WKB}.
We found that the sums converged more quickly if we summed over $n$ first and then
$\ell $.
Convergence of the sums was speeded up by using a Shanks transformation (see appendix D of \cite{howard}).
The disadvantage of the Shanks transformation is that one loses accuracy (typically the accuracy after the transformation
is half that before), but we are nonetheless confident that our final answers are accurate to five significant figures.
The form of the mode sum (\ref{eq:modesum}) will be shown explicitly as a function of $r$ in section \ref{sec:answer}.

\subsection{Numerical Integrals}
\label{sec:integrals}

\begin{figure}[h]
\begin{center}
\includegraphics[width=6cm,angle=270]{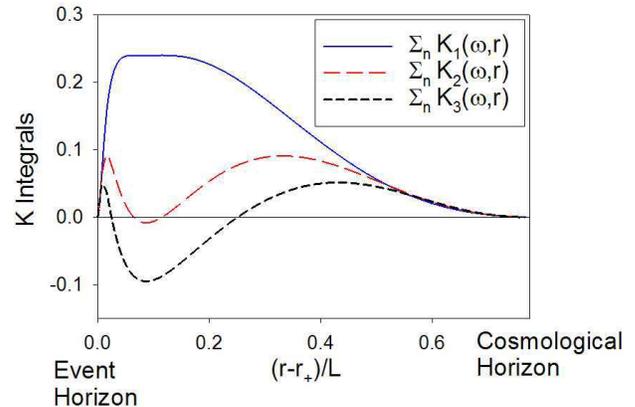}
\end{center}
\caption{
Sums over $n$ of the $K_{i}(\omega ,r )$ integrals (\ref{eq:K1}--\ref{eq:K3})
as functions of $r$.
Note that these are independent of the mass or coupling of the scalar field.
The radial co-ordinate on the horizontal axis is in units of $L$, and we consider the black hole
metric (\ref{eq:metric},\ref{eq:fRNdS}) with parameter $M=Q=0.1L$.
It can be seen that all three sums vanish at the event horizon (left-hand-edge of the plot)
and the cosmological horizon (right-hand-edge of the plot).
This is in agreement with analytic work to be presented in section \ref{sec:analhor}.}
\label{fig:Kintegrals}
\end{figure}

The remainder of $\langle \phi ^{2} \rangle _{\rm {numeric}}$ involves sums of integrals which have to be computed
numerically (\ref{eq:phi2numsplit}):
\begin{equation}
\langle \phi ^{2} \rangle _{\rm {numeric}}^{II} =
\frac {T}{2\pi } \sum _{n=1}^{\infty } \left[  J_{0}(\omega, r) + J_{1}(\omega ,r ) \right] ,
\label{eq:numintbit}
\end{equation}
where $J_{0}(\omega, r)$ is given by (\ref{eq:J0}) and $J_{1}(\omega ,r)$ is given in (\ref{eq:J1}) as
a combination of three integrals, $K_{1}(\omega ,r)$ (\ref{eq:K1}), $K_{2}(\omega ,r)$ (\ref{eq:K2}) and
$K_{3}(\omega ,r )$ (\ref{eq:K3}).
Each of these integrals depends on $a$ (\ref{eq:adef}), which in turn depends on $\omega = 2\pi n T$, for
$n=1,\ldots $ (note that the numerical integrals are only relevant for $\omega \neq 0$, as the modes for
$\omega = 0$ were considered separately in section \ref{sec:WKB}).
For each value of $n$, the integrals are straightforward to compute, and then summed over as many $n$ as required
for convergence.
The summands are all $O(n^{-3})$ as $n \rightarrow \infty $ so convergence is relatively rapid.
As an example of our results, we plot in Fig. \ref{fig:Kintegrals}
\begin{equation}
\sum _{n=1}^{\infty } K_{i}(\omega ,r)
\end{equation}
for $i=1,2,3$, showing how these sums vary with $r$.

From (\ref{eq:K1}--\ref{eq:K3}), these sums are independent of the mass or coupling of the
scalar field, but do depend on the metric of the black hole through the function $f$ (\ref{eq:fRNdS}).
It can be clearly seen in Fig. \ref{fig:Kintegrals} that all three sums vanish at both the
event and cosmological horizons.
We will show analytically in section \ref{sec:analhor} that this is always the case, independent of the
temperature of the thermal state under consideration.

\subsection{Final Answers}
\label{sec:answer}

The final answers for $\langle \phi ^{2} \rangle _{\rm {ren}}$ are now computed by combining the mode
sum (\ref{eq:modesum}) with the numerical integral contribution (\ref{eq:numintbit}) to give
$\langle \phi ^{2} \rangle _{\rm {numeric}}$ (\ref{eq:phi2ren}), and then adding this to
$\langle \phi ^{2} \rangle _{\rm {analytic}}$ (\ref{eq:phi2analytic}).

\begin{figure}[h]
\begin{center}
\includegraphics[width=6cm,angle=270]{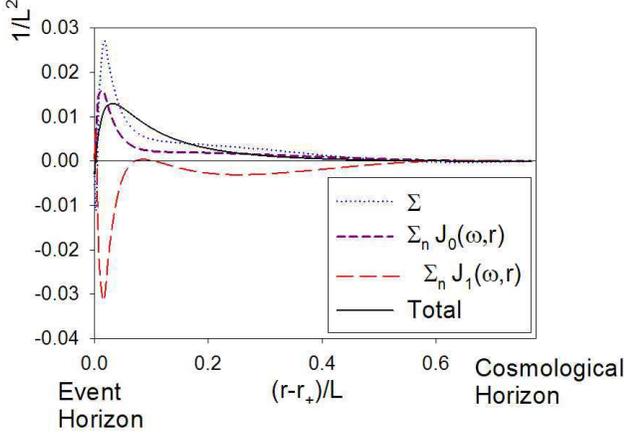}
\end{center}
\caption{Contributions to $\langle \phi ^{2} \rangle _{\rm {numeric}}$ (\ref{eq:phi2ren}) from the
mode sum $\Sigma $ (\ref{eq:modesum}) and the sums over the numerical integrals (\ref{eq:numintsums}),
as functions of $r$.
Here, the quantum scalar field is massless and conformally coupled.
The radial co-ordinate on the horizontal axis is in units of $L$,
and the black hole metric (\ref{eq:metric},\ref{eq:fRNdS}) has parameter $M=Q=0.1L$.
All quantities appear to be regular near the cosmological horizon.
Their behaviour near the event horizon is shown in close-up in Fig. \ref{fig:phi2numm0close}.
}
\label{fig:phi2numm0}
\end{figure}

\begin{figure}[h]
\begin{center}
\includegraphics[width=5cm,angle=270]{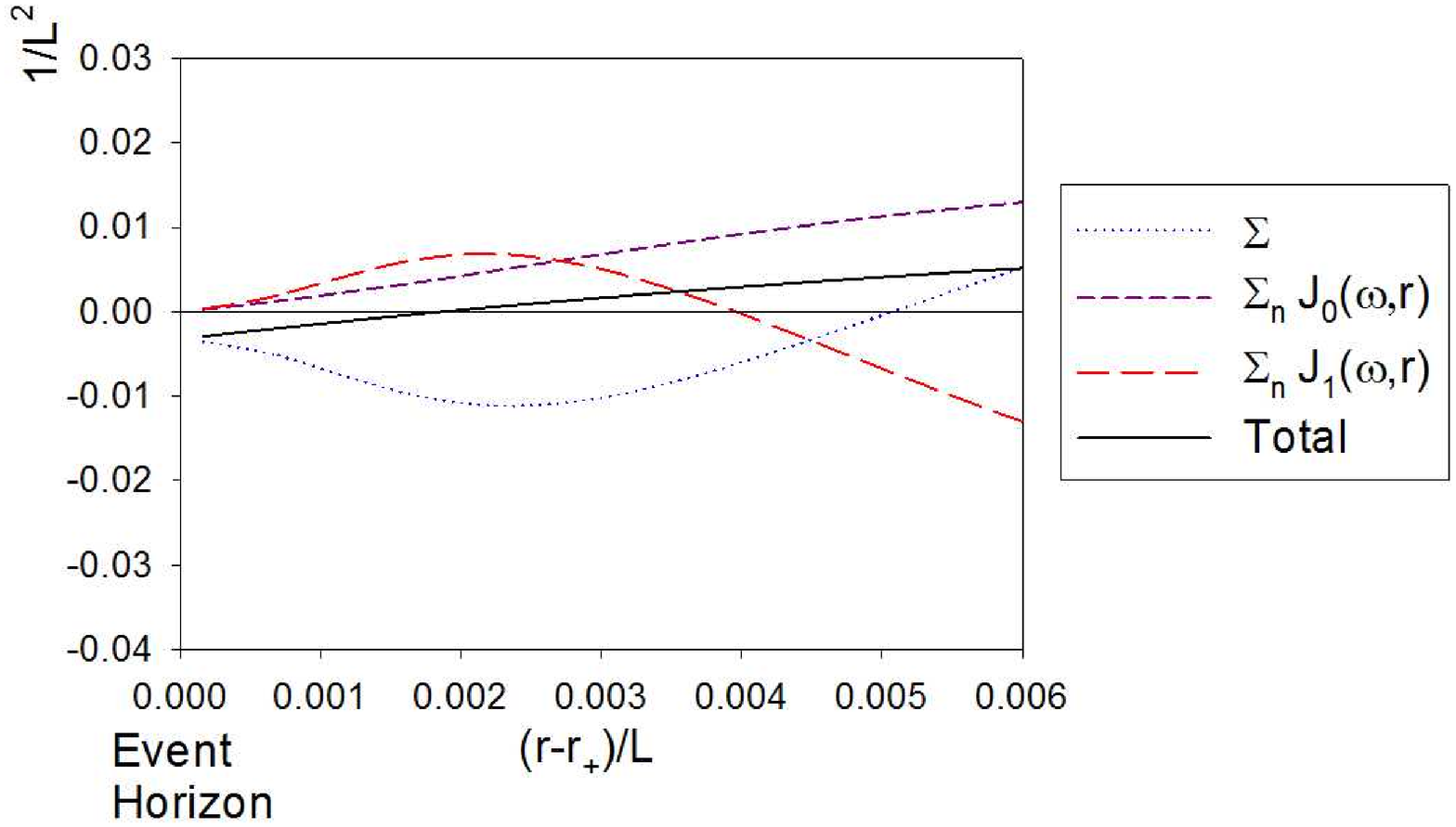}
\end{center}
\caption{
Close-up, near the event horizon, of the contributions (shown in Fig. \ref{fig:phi2numm0}) to
$\langle \phi ^{2} \rangle _{\rm {numeric}}$ (\ref{eq:phi2ren}) from the
mode sum $\Sigma $ (\ref{eq:modesum}) and the sums over the numerical integrals (\ref{eq:numintsums}),
when the quantum scalar field is massless and conformally coupled.
It is clear that the contributions from the sums over the numerical integrals (\ref{eq:numintsums})
vanish at the event horizon.
Near the horizon the mode sum $\Sigma $ (\ref{eq:modesum}) is the dominant contribution and
this appears to remain finite as the horizon is approached.}
\label{fig:phi2numm0close}
\end{figure}

\begin{figure}[h]
\begin{center}
\includegraphics[width=6cm,angle=270]{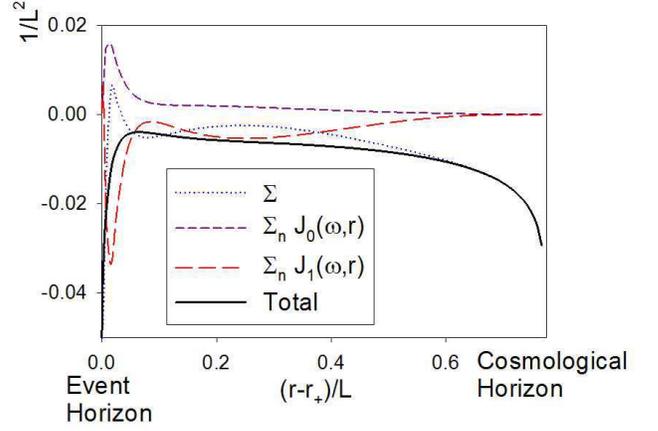}
\end{center}
\caption{
Contributions to $\langle \phi ^{2} \rangle _{\rm {numeric}}$ (\ref{eq:phi2ren}) from the
mode sum $\Sigma $ (\ref{eq:modesum}) and the sums over the numerical integrals (\ref{eq:numintsums}),
as functions of $r$.
Here, the quantum scalar field is conformally coupled but has mass $m=L$.
The radial co-ordinate on the horizontal axis is in units of $L$,
and the black hole metric (\ref{eq:metric},\ref{eq:fRNdS}) has parameter $M=Q=0.1L$.
The contributions from the sums over the numerical integrals appear to be regular near the cosmological
horizon, but the mode sum $\Sigma $ (\ref{eq:modesum}) is diverging as the cosmological horizon is
approached.
The behaviour of these quantities near the event horizon is shown in close-up in Fig. \ref{fig:phi2numm1close}.
}
\label{fig:phi2numm1}
\end{figure}

\begin{figure}[h]
\begin{center}
\includegraphics[width=5cm,angle=270]{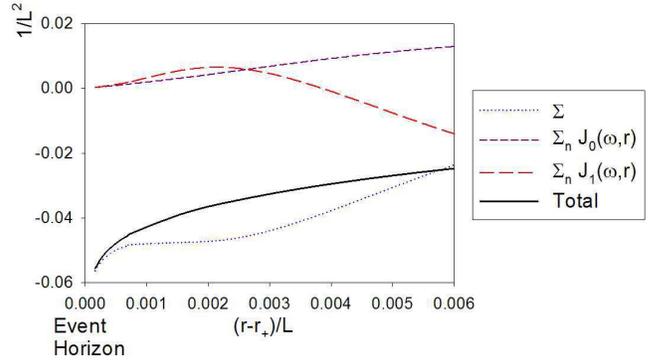}
\end{center}
\caption{
Close-up, near the event horizon, of the contributions (shown in Fig. \ref{fig:phi2numm1}) to
$\langle \phi ^{2} \rangle _{\rm {numeric}}$ (\ref{eq:phi2ren}) from the
mode sum $\Sigma $ (\ref{eq:modesum}) and the sums over the numerical integrals (\ref{eq:numintsums}),
when the quantum scalar field is conformally coupled and has mass $m=L$.
It is clear that the contributions from the sums over the numerical integrals (\ref{eq:numintsums})
vanish at the event horizon.
Near the horizon the mode sum $\Sigma $ (\ref{eq:modesum}) is the dominant contribution and
this appears to be diverging as the horizon is approached.
}
\label{fig:phi2numm1close}
\end{figure}

We begin this section by examining the contributions to $\langle \phi ^{2} \rangle _{\rm {numeric}}$,
namely the mode sum (\ref{eq:modesum}), and the two sums over numerical integrals
\begin{equation}
\frac {T}{2\pi }\sum _{n=1}^{\infty } J_{0}(\omega ,r ), \qquad
\frac {T}{2\pi }\sum _{n=1}^{\infty } J_{1}(\omega ,r ).
\label{eq:numintsums}
\end{equation}
These are shown in Figs. \ref{fig:phi2numm0}--\ref{fig:phi2numm1close} for a lukewarm black hole
with $M=Q=0.1L$.
The quantity labeled `Total' in Figs. \ref{fig:phi2numm0}--\ref{fig:phi2numm1close}
is $\langle \phi ^{2} \rangle _{\rm {numeric}}$.


In Figs. \ref{fig:phi2numm0} and \ref{fig:phi2numm0close} we consider a massless, conformally
coupled, scalar field, while in Figs. \ref{fig:phi2numm1} and \ref{fig:phi2numm1close} the
field is still conformally coupled but has mass $m=L$.
We have chosen a large value of the mass so that the differences between this and the massless case
are very clear.

Examining first the massless case, from Fig. \ref{fig:phi2numm0} we can see that all three contributions
to $\langle \phi ^{2} \rangle _{\rm {numeric}}$ have peaks near the event horizon of the black hole, and that all
three parts become very small at the cosmological horizon.
The behaviour near the event horizon is clearer in the close up in Fig. \ref{fig:phi2numm0close}.
From Fig. \ref{fig:phi2numm0close} it can be seen that the contributions coming from sums over the numerical integrals
(\ref{eq:numintsums}) tend to zero as the horizon is approached, and the dominant contribution to
$\langle \phi ^{2} \rangle _{\rm {numeric}}$ comes from the mode sum $\Sigma $ (\ref{eq:modesum}).
Furthermore, this appears to remain finite as the horizon is approached.

When the scalar field is no longer massless, we see the difference in behaviour clearly in Fig. \ref{fig:phi2numm1}.
Near the cosmological horizon, the contributions to $\langle \phi ^{2} \rangle _{\rm {numeric}}$ from the
sums over the numerical integrals (\ref{eq:numintsums}) still vanish as the horizon is approached, but now
the mode sum $\Sigma $ (\ref{eq:modesum}) (and therefore $\langle \phi ^{2} \rangle _{\rm {numeric}}$) diverges.
The close-up near the event horizon in Fig. \ref{fig:phi2numm1close} confirms that this also happens near
the black hole event horizon.

We will show in section \ref{sec:regular} that the contributions to $\langle \phi ^{2} \rangle _{\rm {numeric}}$
from the sums over numerical integrals always vanish at the horizons,
whatever the mass or coupling of the scalar field, so that the mode sum $\Sigma $ is
indeed the dominant contribution to $\langle \phi ^{2} \rangle _{\rm {numeric}}$ near the horizons.
Furthermore, we will show that $\Sigma $ diverges near a horizon unless the quantum scalar field is both
massless and conformally coupled.

\begin{figure}[h]
\begin{center}
\includegraphics[width=6cm,angle=270]{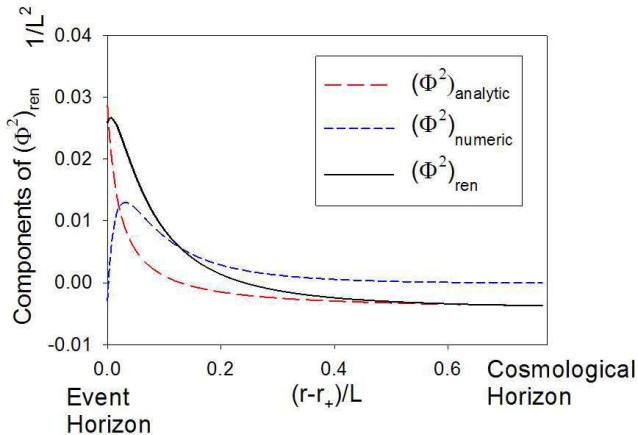}
\end{center}
\caption{
The vacuum polarization $\langle \phi ^{2} \rangle _{\rm {ren}}$ and its
components $\langle \phi ^{2} \rangle _{\rm {analytic}}$
(\ref{eq:phi2analytic}) and $\langle \phi ^{2} \rangle _{\rm {numeric}}$ (\ref{eq:phi2ren})
(all in units of $L^{-2}$).
Here, the quantum scalar field is massless and conformally coupled.
The radial co-ordinate on the horizontal axis is in units of $L$,
and the black hole metric (\ref{eq:metric},\ref{eq:fRNdS}) has parameter $M=Q=0.1L$.
All quantities appear to be finite near both the event and cosmological horizons.
Near the event and cosmological horizons, $\langle \phi ^{2} \rangle _{\rm {analytic}}$
is the dominant contribution to $\langle \phi ^{2} \rangle _{\rm {ren}}$.
}
\label{fig:phi2renm0}
\end{figure}

\begin{figure}[h]
\begin{center}
\includegraphics[width=5cm,angle=270]{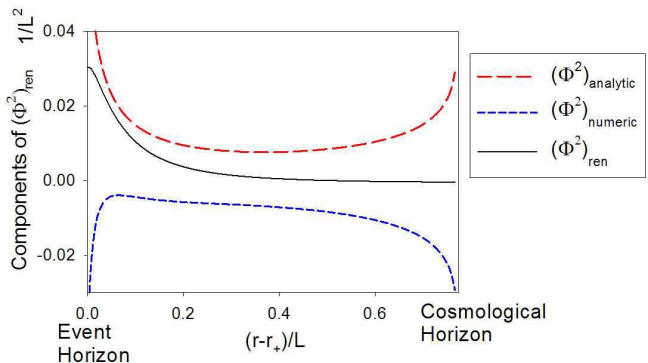}
\end{center}
\caption{
The vacuum polarization $\langle \phi ^{2} \rangle _{\rm {ren}}$ and its
components $\langle \phi ^{2} \rangle _{\rm {analytic}}$
(\ref{eq:phi2analytic}) and $\langle \phi ^{2} \rangle _{\rm {numeric}}$ (\ref{eq:phi2ren})
(all in units of $L^{-2}$).
Here, the quantum scalar field is conformally coupled and has mass $m=L$.
The radial co-ordinate on the horizontal axis is in units of $L$,
and the black hole metric (\ref{eq:metric},\ref{eq:fRNdS}) has parameter $M=Q=0.1L$.
It can be clearly seen that both $\langle \phi ^{2} \rangle _{\rm {analytic}}$
and $\langle \phi ^{2} \rangle _{\rm {numeric}}$ diverge near the event and cosmological
horizons, but with opposite signs.
}
\label{fig:phi2renm1}
\end{figure}

We now turn to the final result, namely $\langle \phi ^{2} \rangle _{\rm {ren}}$.
In Figs. \ref{fig:phi2renm0}--\ref{fig:phi2renm1} we plot $\langle \phi ^{2} \rangle _{\rm {analytic}}$
(\ref{eq:phi2analytic}), $\langle \phi ^{2} \rangle _{\rm {numeric}}$ (\ref{eq:phi2ren}) and
their sum $\langle \phi ^{2} \rangle _{\rm {ren}}$ for a massless, conformally coupled scalar field, and
a conformally coupled scalar field with mass $m=L$.

%\begin{figure}
%\begin{center}
%\includegraphics[width=6cm,angle=270]{phi2ren_m0_close.eps}
%\end{center}
%\caption{
%Close-up, near the horizon, of the vacuum polarization $\langle \phi ^{2} \rangle _{\rm {ren}}$ and its
%components $\langle \phi ^{2} \rangle _{\rm {analytic}}$
%(\ref{eq:phi2analytic}) and $\langle \phi ^{2} \rangle _{\rm {numeric}}$ (\ref{eq:phi2ren})
%(all in units of $L^{-2}$), when the quantum scalar field is massless and conformally coupled.
%It is clear that all quantities are finite, and that $\langle \phi ^{2} \rangle _{\rm {analytic}}$
%is the dominant contribution to $\langle \phi ^{2} \rangle _{\rm {ren}}$.
%}
%\label{fig:phi2renm0close}
%\end{figure}
%\begin{figure}
%\begin{center}
%\includegraphics[width=6cm,angle=270]{phi2ren_m1_close.eps}
%\end{center}
%\caption{$m=1$.
%For these mode functions, we have taken the scalar field to be massless and conformally coupled,
%and the black hole metric (\ref{eq:metric},\ref{eq:fRNdS}) has parameter $M=Q=0.1L$.
%The radial co-ordinate on the horizontal axis is in units of $L$.}
%\label{fig:phi2renm1close}
%\end{figure}

When the scalar field is massless and conformally coupled, from Fig. \ref{fig:phi2renm0} it is
clear that $\langle \phi ^{2} \rangle _{\rm {analytic}}$,  $\langle \phi ^{2} \rangle _{\rm {numeric}}$
and $\langle \phi ^{2} \rangle _{\rm {ren}}$ are regular at both the event and cosmological horizons.
The total $\langle \phi ^{2} \rangle _{\rm {ren}}$ is greatest near the event horizon, and slightly negative
near the cosmological horizon.
The contribution $\langle \phi ^{2} \rangle _{\rm {analytic}}$ has the same qualitative behaviour as the
total $\langle \phi ^{2} \rangle _{\rm {ren}}$ but does not represent more than a qualitative approximation
to the exact quantity.

When the quantum scalar field is not massless, from Fig. \ref{fig:phi2renm1} we can see that both
$\langle \phi ^{2} \rangle _{\rm {analytic}}$ and  $\langle \phi ^{2} \rangle _{\rm {numeric}}$
diverge as either the event or cosmological horizons are approached, albeit with opposite signs.
From our numerical results, it looks like these divergences cancel to give a finite
$\langle \phi ^{2} \rangle _{\rm {ren}}$.
However, finiteness of this quantity cannot be proven by numerical calculations alone.
Therefore, in the next section, we study $\langle \phi ^{2} \rangle _{\rm {ren}}$
analytically.

\section{Regularity on the Horizons}
\label{sec:regular}

Our numerical results, computed in the previous section, indicate that $\langle \phi ^{2} \rangle _{\rm {ren}}$
is finite on both the event and cosmological horizons, for the thermal state at the same temperature
as both horizons.
In this section we shall show analytically that this is indeed the case.
We shall consider a general, non-extremal horizon at $r=r_{0}$, and show that
$\langle \phi ^{2} \rangle _{\rm {ren}}$ is regular at that horizon for a thermal state
at the same temperature as the horizon.

\subsection{Analytic Part and Numerical Integrals}
\label{sec:analhor}

We consider first the behaviour of $\langle \phi ^{2} \rangle _{\rm {analytic}}$ at a horizon $r=r_{0}$.
In order to analyze the regularity at a horizon of $\langle \phi ^{2} \rangle _{\rm {analytic}}$
(\ref{eq:phi2analytic}), we split it into two parts:
\begin{equation}
\langle \phi ^{2} \rangle _{\rm {analytic}}  =
\langle \phi ^{2} \rangle _{\rm {analytic}}^{I} + \langle \phi ^{2} \rangle _{\rm {analytic}}^{II};
\end{equation}
where
\begin{eqnarray}
\langle \phi ^{2} \rangle _{\rm {analytic}}^{I} & = &
 \frac {m^{2}}{16\pi ^{2}}
-\frac {1}{192\pi ^{2}f}\left( \frac {df}{dr} \right) ^{2}
+ \frac {1}{96\pi ^{2}} \frac {d^{2}f}{dr^{2}}
\nonumber \\ & &
+ \frac {1}{48\pi ^{2}r} \frac {df}{dr} + \frac {\kappa ^{2}}{48 \pi ^{2}f} ;
\nonumber \\
 \langle \phi ^{2} \rangle _{\rm {analytic}}^{II} & = &
- \frac {1}{8\pi ^{2}} \left[ m^{2} + \left( \xi - \frac {1}{6} \right) R \right]
\nonumber \\ & & \times
\left[ C +\frac {1}{2} \ln \left( \frac {\mu ^{2} f }{4\kappa ^{2}} \right) \right] .
\label{eq:phi2analyticII}
\end{eqnarray}
At first sight it looks like $\langle \phi ^{2} \rangle _{\rm {analytic}}^{I}$ is divergent at a horizon
where $f=0$, but in fact it is straightforward to show that if $\kappa $ is chosen to be
$\frac {1}{2}\frac {df}{dr}$ at the horizon, then $\langle \phi ^{2} \rangle _{\rm {analytic}}^{I}$
is regular there.
This is in accordance with expectations: if the temperature $T=2\pi \kappa $
of the state matches the natural temperature
of the horizon we are considering, then $\langle \phi ^{2} \rangle _{\rm {analytic}}^{I}$ is regular at
that horizon.
In particular, for the lukewarm black holes, we have that $\langle \phi ^{2} \rangle _{\rm {analytic}}^{I}$
is regular at both the event and cosmological horizons.

On the other hand, it is clear that $\langle \phi ^{2} \rangle _{\rm {analytic}}^{II}$ has a logarithmic divergence
as $f\rightarrow 0$ unless both $m^{2}=0$ and $\xi = 1/6$, that is, the field is massless and conformally
coupled.
We will show in the next subsection that this divergence cancels with a corresponding divergence in the
mode sum $\Sigma $ (\ref{eq:modesum}), to give an overall $\langle \phi ^{2} \rangle _{\rm {ren}}$ which is regular.

Before that, we examine next the other contribution to $\langle \phi ^{2} \rangle _{\rm {numeric}}$, namely
\begin{equation}
\langle \phi ^{2} \rangle _{\rm {numeric}}^{II} =
\frac {T}{2\pi } \sum _{n=1}^{\infty } \left[  J_{0}(\omega, r) + J_{1}(\omega ,r ) \right] ,
\label{eq:phi2numericII}
\end{equation}
where $J_{0}(\omega ,r)$ and $J_{1}(\omega ,r)$ are given, respectively, by (\ref{eq:J0},\ref{eq:J1}),
with $J_{1}(\omega, r)$ in turn determined by the integrals in (\ref{eq:K1}--\ref{eq:K3}).
It is straightforward to show that, for fixed $\omega $ and small $f$, the integrals $J_{0}(\omega ,r)$,
$K_{1}(\omega ,r)$, $K_{2}(\omega ,r)$ and $K_{3}(\omega ,r)$ behave as follows:
\begin{eqnarray}
J_{0}(\omega, r) & = &
\frac {7f}{1920\omega ^{3}r_{0}^{4}} + O(f^{2});
\nonumber \\
K_{1}(\omega ,r ) & = &
\frac {f^{\frac {3}{2}}}{48 \omega ^{3} r^{3}} + \frac {21f^{\frac {5}{2}}}{3840\omega ^{5}r_{0}^{5}}
+ O(f^{\frac {7}{2}});
\nonumber \\
K_{2} (\omega ,r) & = &
\frac {f^{\frac {3}{2}}}{48 \omega ^{3} r^{3}} + \frac {7f^{\frac {5}{2}}}{768\omega ^{5}r_{0}^{5}}
+ O(f^{\frac {7}{2}});
\nonumber \\
K_{3} (\omega ,r) & = &
\frac {f^{\frac {3}{2}}}{48 \omega ^{3} r^{3}} + \frac {49f^{\frac {5}{2}}}{3840\omega ^{5}r_{0}^{5}}
+ O(f^{\frac {7}{2}});
\label{eq:Kinthor}
\end{eqnarray}
where $r=r_{0}$ is the location of the horizon where $f=0$.
We should comment that the first terms in the expansions of $K_{1}(\omega ,r )$, $K_{2}(\omega ,r )$
and $K_{3}(\omega ,r)$ do have $r$ and not $r_{0}$ in the denominator.
Therefore they include some second order terms as well as leading order terms.
Writing the expansions in this way simply makes the expressions shorter.
From (\ref{eq:Kinthor}), it is clear that the sums over the $K_{i}(\omega ,r)$ should vanish at
the horizon, regardless of the temperature $T$ of the state under consideration or the mass or coupling
of the scalar field.
This is in agreement with our numerical results in Fig. \ref{fig:Kintegrals}.

Combining the $K_{i}(\omega ,r)$ integrals to form $J_{1}(\omega ,r)$ using (\ref{eq:J1}), we find that
the leading order contributions from the $K_{i}(\omega ,r)$ integrals cancel and
$J_{1}(\omega ,r)$ has the following behaviour:
\begin{equation}
J_{1} (\omega ,r) = \frac {7f}{7680\omega ^{5} r_{0}^{4}} \left( \frac {df}{dr} \right) ^{2} + O(f^{2}).
\end{equation}
Recalling that $\omega = 2n\pi T$, $n\neq0 $,
we therefore see that both $J_{0}(\omega ,r )$ and $J_{1}(\omega ,r )$ tend to zero uniformly in $n$
as $f\rightarrow 0$, and therefore the contributions to
$\langle \phi ^{2} \rangle _{\rm {numeric}}$ from the numerical integrals (\ref{eq:phi2numericII})
vanish at both the event and cosmological
horizons.
This is independent of the (non-zero) temperature of the state under consideration, and the mass and curvature
coupling of the scalar field, and is in agreement with our numerical results
in Figs. \ref{fig:phi2numm0}--\ref{fig:phi2numm1close}.

Therefore the only parts of $\langle \phi ^{2} \rangle _{\rm {ren}}$ which need further investigation
are $\langle \phi ^{2} \rangle _{\rm {analytic}}^{II}$ (\ref{eq:phi2analyticII}), which has a logarithmic
divergence as $r\rightarrow r_{0}$, and the mode sum (\ref{eq:modesum}) contribution to
$\langle \phi ^{2} \rangle _{\rm {numeric}}$.
If $\langle \phi ^{2} \rangle _{\rm {ren}}$ is to be regular at the horizon, it must be the case that
the mode sum $\Sigma $ (\ref{eq:modesum}) also has a logarithmic divergence, and, furthermore, this
must exactly cancel the logarithmic divergence in $\langle \phi ^{2} \rangle _{\rm {analytic}}^{II}$.
We next establish that this is the case using a careful analysis of the mode functions.

\subsection{Mode Sums}
\label{sec:modehor}

The behaviour of the mode sum $\Sigma $ (\ref{eq:modesum}) near a horizon is by far the most difficult
part of $\langle \phi ^{2} \rangle _{\rm {ren}}$ to analyze.
The reason for this is that the sums over $\ell $ are not uniformly convergent: more and more $\ell $'s are
needed before convergence is reached as we get closer and closer to a horizon.
This is due to the fact that the WKB approximants $\beta _{0\omega \ell }(r)$ (\ref{eq:beta0})
and $\beta _{1\omega \ell}(r)$ (\ref{eq:beta1}) do not depend on $\ell $ when $f=0$, as in this case
$\chi _{\omega \ell }(r)$ (\ref{eq:chi}) reduces to
\begin{equation}
\chi _{\omega \ell }(r) = \omega r^{2}.
\end{equation}
Furthermore, the WKB approximants do not correctly reproduce the $\ln [(r-r_{0})/r_{0}]$ behaviour
seen in the field modes (\ref{eq:Frobeniusmodes}) near the horizon.

Suppose we are considering a horizon on which the mode functions $p_{\omega \ell }(r)$ are regular and the
$q_{\omega \ell }(r)$ diverge.
The analysis below will be equally valid if we are on a horizon where the $p_{\omega \ell }(r)$ diverge
and the $q_{\omega \ell }(r)$ are regular, but with the $p$'s and $q$'s swapped over.
Then near the horizon, the mode functions $p_{\omega \ell }(r)$ and $q_{\omega \ell }(r)$ will have the
following behaviour (\ref{eq:Frobeniusmodes}):
\begin{eqnarray}
p_{\omega \ell } (r) & = & (r-r_{0})^{\frac {n}{2}} + O(r - r_{0})^{1+\frac {n}{2}} ;
\nonumber \\
q_{\omega \ell } (r) & = & (r-r_{0})^{-\frac {n}{2}} + O(r - r_{0})^{1-\frac {n}{2}}
\nonumber \\ & &
+ {\cal {K}}_{\omega \ell } (r-r_{0})^{\frac {n}{2}}  \ln \left( \frac {r-r_{0}}{r_{0}} \right)
\nonumber \\ & &
+ O\left[(r-r_{0})^{1+\frac {n}{2}} \ln \left( \frac {r-r_{0}}{r_{0}} \right) \right] ;
\end{eqnarray}
for some constant ${\cal {K}}_{\omega \ell }$, where $\omega = 2\pi n T$.
Therefore, near the horizon, we have
\begin{eqnarray}
C_{\omega \ell } p_{\omega \ell }(r) q_{\omega \ell }(r)  & \sim &
C_{\omega \ell }{\cal {K}}_{\omega \ell } (r-r_{0})^{n}
\ln \left( \frac {r-r_{0}}{r_{0}} \right)
\nonumber \\  & &
+ {\mbox {finite terms.}}
\label{eq:Cpqhor}
\end{eqnarray}
From (\ref{eq:Cpqhor}), it is clear that the only contribution to the mode sum $\Sigma $ which
is important near the horizon is that for $n=0=\omega $, since this contains a $\ln [(r-r_{0})/r_{0}]$ term
which diverges at the horizon.
For $n>0$, the $(r-r_{0})^{n}\ln [(r-r_{0})/r_{0}]$ term is subleading compared to the finite terms.
The finite terms in (\ref{eq:Cpqhor})
for $n>0$ are described well by the conventional WKB approximation described in section \ref{sec:WKB}
and their contribution to the total mode sum is finite as the horizon is approached \cite{tomimatsu}.

To analyze the behaviour of the mode sum $\Sigma $,
what is therefore required is an approximation for the mode functions with $n= \omega = 0$,
near the horizon, which is {\em {uniformly}} valid in $\ell $.
Such an approximation was first developed by Candelas \cite{candelas} for Schwarzschild black holes,
and subsequently extended in Ref. \cite{tomimatsu} for the Reissner-Nordstr\"om black hole.
Here we generalize the approach of Ref. \cite{tomimatsu} for general metric function $f(r)$.
We focus on the case of a horizon at $r=r_{0}$, where $p_{0 \ell }(r)$ is regular and $q_{0 \ell }(r)$ diverges,
but the analysis proceeds similarly for a horizon where $q_{0\ell }(r)$ is regular and
$p_{0\ell }(r)$ diverges.

We begin by writing the mode functions in terms of modified Bessel functions \cite{tomimatsu}:
\begin{eqnarray}
p_{0\ell }(r) & = &
\left( \frac{\Gamma (r)}{r^{2} \Omega (r)} \right) ^{\frac {1}{2}} I_{0} (\Gamma (r) );
\nonumber \\
q_{0\ell }(r) & = &
\left( \frac{\Gamma (r)}{r^{2} \Omega (r)} \right) ^{\frac {1}{2}} K_{0} (\Gamma (r) );
\label{eq:modeBessel}
\end{eqnarray}
where $\Gamma (r)$ and $\Omega (r)$ are functions which we shall define very shortly.
The mode functions (\ref{eq:modeBessel}) are required to satisfy the differential equation
(\ref{eq:modes}) and the normalization condition (\ref{eq:Wronskian}).
Considering the normalization condition (\ref{eq:Wronskian}) first, this is identically satisfied with
$C_{0\ell }=1$ if $\Gamma (r)$ is defined to be
\begin{equation}
\Gamma (r) = \int _{r'=r_{0}}^{r} \frac {\Omega (r')}{f(r')} \, dr' ,
\label{eq:Gamma}
\end{equation}
whatever the (as yet undetermined) function $\Omega (r)$.
With this definition of $\Gamma (r)$, substituting (\ref{eq:modeBessel}) into (\ref{eq:modes}) gives
a differential equation for $\Omega (r)$ which can be found in \cite{tomimatsu}.
Following \cite{tomimatsu}, the function $\Omega (r)$ is defined to be
\begin{equation}
\Omega (r) = y(r) {\sqrt {f(r)}},
\label{eq:Omega}
\end{equation}
where $y(r)$ is a function which is regular at the horizon.
To analyze the behaviour of the mode functions near the horizon, we expand $f(r)$ and $y(r)$
near the horizon as follows:
\begin{eqnarray}
f(r) & = & x\left[ f_{1} + xf_{2} + O(x^{2}) \right] ;
\nonumber \\
y(r) & = & B \left[ 1 + xy_{1} + x^{2} y_{2} + O(x^{3}) \right] ;
\label{eq:horexp}
\end{eqnarray}
where $f_{i}$, $B$ and $y_{i}$ are constants and $x=r-r_{0}$.
The fact that this expansion should be {\em {uniform}} in $\ell $ means that the constants
$y_{i}$ must be bounded as $\ell \rightarrow \infty $.
The expansions (\ref{eq:horexp}) lead to the following behaviour near the horizon:
\begin{eqnarray}
\Gamma (r) & = & \frac {2B}{r_{0}} {\sqrt {\frac {x}{f_{1}}}} + O (x^{\frac {3}{2}}) ;
\nonumber \\
\left( \frac {\Gamma (r)}{r^{2} \Omega (r)} \right) ^{\frac {1}{2}}
& = &
{\sqrt {\frac {2}{f_{1}r_{0}^{2}}}} + O(x).
\label{eq:GOhor}
\end{eqnarray}

The expansions (\ref{eq:horexp}) are substituted, in turn,
into (\ref{eq:Omega}), then (\ref{eq:Gamma}) and finally (\ref{eq:modeBessel})
to give the form of the mode functions near the horizon.
These are then, in turn, substituted into the mode equation (\ref{eq:modes}), which is then expanded in powers
of $x$.
The lowest order term in (\ref{eq:modes}) gives the following expression for the constant $B$:
\begin{eqnarray}
B^{2}
 & = &  \ell \left( \ell + 1 \right) +  \frac {1}{3} r_{0}f_{1} + \frac {1}{3} r_{0}^{2} f_{1} y_{1}
+ \frac {1}{3} + m^{2} r_{0}^{2}
\nonumber \\ & &
+ \left( \xi - \frac {1}{6} \right) R_{0} r_{0}^{2},
\label{eq:B}
\end{eqnarray}
where $R_{0}=R(r_{0})$ is the value of the Ricci scalar on the horizon.
The expression (\ref{eq:B}) generalizes that found in \cite{tomimatsu} for the Reissner-Nordstr\"om case.
As in \cite{tomimatsu}, the next-to-leading order term in the mode equation (\ref{eq:modes}) gives a complicated
expression involving $y_{1}$ and $y_{2}$.
Using the fact that both $y_{1}$ and $y_{2}$ must remain bounded as $\ell \rightarrow \infty $, we obtain
\begin{equation}
y_{1} = - \frac {1}{r_{0}} + O(\ell ^{-2}) \qquad {\mbox {as $\ell \rightarrow \infty $}},
\end{equation}
which means that, as $\ell \rightarrow \infty $, we have
\begin{equation}
B^{2} = \ell \left( \ell + 1 \right) + \frac {1}{3} + m^{2} r_{0}^{2}
+ \left( \xi - \frac {1}{6} \right) R_{0} r_{0}^{2}
+ O(\ell ^{-2}).
\label{eq:BlargeL}
\end{equation}

We may now use the form of the mode functions (\ref{eq:modeBessel}) to study the contribution of the
$\omega = 0$ modes to the mode sum $\Sigma $ (\ref{eq:modesum}) near the horizon.
In other words, we wish to analyze
\begin{eqnarray}
\Sigma _{0} & = &
\frac {T}{4\pi } \sum _{\ell = 0}^{\infty } \left[ \left( 2 \ell + 1 \right)
C_{0 \ell } p_{0 \ell }(r) q_{0 \ell }(r)
- \frac {1}{rf^{\frac {1}{2}}} \right]
\nonumber \\ & = &
\frac {T}{4\pi } \sum _{\ell = 0}^{\infty } \left[ \left( 2 \ell + 1 \right)
\left( \frac{\Gamma (r)}{r^{2} \Omega (r)} \right) I_{0}(\Gamma (r)) K_{0}(\Gamma (r))
\right. \nonumber \\ & & \left.
- \frac {1}{rf^{\frac {1}{2}}} \right] .
\label{eq:Sigma0}
\end{eqnarray}
In section \ref{sec:WKB}, we added and subtracted $\sum _{\ell =0}^{\infty } \beta _{10\ell }(r)$
to this sum to give a sum which was more readily computed numerically.
We do not need to do this here as the sum is convergent.
In addition, $\beta _{10\ell }(r)$ diverges as $r\rightarrow r_{0}$, so that the sum (\ref{eq:Sigma0})
is more straightforward to analyze, as it does not include these additional divergences (which do, however, cancel).

Using the Watson-Sommerfeld formula (\ref{eq:watson}), we convert the sum (\ref{eq:Sigma0})
over $\ell $ into two integrals:
\begin{equation}
\Sigma _{0}  =
\frac {T}{4\pi } \left[
\Sigma _{0}^{I} + \Sigma _{0}^{II} \right] ,
\end{equation}
where
\begin{eqnarray}
\Sigma _{0}^{I} & = &
\int _{0}^{\infty } d\lambda \left[ 2\lambda
\left( \frac{\Gamma (r)}{r^{2} \Omega (r)} \right) I_{0}(\Gamma (r)) K_{0}(\Gamma (r))
\right. \nonumber \\ & & \left.
- \frac {1}{rf^{\frac {1}{2}}} \right] ;
\label{eq:Sigma0one}
\\
\Sigma _{0}^{II} & = &
- \Re \left\{ i \int _{0}^{\infty } d\lambda \frac {2}{1+e^{2\pi \lambda }} \right.
\nonumber \\ & & \times \left.
\left[ 2 i\lambda
\left( \frac{\Gamma (r)}{r^{2} \Omega (r)} \right) I_{0}(\Gamma (r)) K_{0}(\Gamma (r))
- \frac {1}{rf^{\frac {1}{2}}} \right] \right\} .
\nonumber \\
\label{eq:Sigma0two}
\end{eqnarray}
In (\ref{eq:Sigma0one}) we have $\ell = \lambda - 1/2$, while in (\ref{eq:Sigma0two}) we have
$\ell = i\lambda - 1/2$.
In both cases the functions $\Gamma (r)$ and $\Omega (r)$ depend on $\ell $.

We now use the expansions (\ref{eq:GOhor}) for the functions $\Gamma (r)$ and $\Omega (r)$ near the
horizon, which are uniform in $\ell $.
It turns out that $\Sigma _{0}^{II}$ is the more straightforward to work with, so we examine this first.
Using (\ref{eq:GOhor}), the leading order behaviour of $\Sigma _{0}^{II}$ near the horizon is
given by
\begin{eqnarray}
\Sigma _{0}^{II} & = &
\Re \left\{
\int _{0}^{\infty } d\lambda
\frac {8\lambda }{r_{0}^{2}f_{1} \left( 1+ e^{2\pi \lambda } \right)}
\right. \nonumber \\ & & \left. \times
 I_{0}\left( \frac {2B}{r_{0}} {\sqrt {\frac {x}{f_{1}}}} \right)
 K_{0}\left( \frac {2B}{r_{0}} {\sqrt {\frac {x}{f_{1}}}} \right) \right\}
 \nonumber \\ & &
 + O(x\ln [x/r_{0}]),
\end{eqnarray}
where $B$ depends on $\lambda $.
In this case we can use the approximate behaviour of the modified Bessel functions for small $x$ \cite{abramowitz}:
\begin{eqnarray}
I_{0}(x) & = & 1 + O(x^{2});
\nonumber \\
 K_{0}(x) & = & -C - \ln \left( \frac {x}{2}\right)  + O(x^{2}\ln x) ;
 \label{eq:I0K0hor}
\end{eqnarray}
where $C$ is Euler's constant.
We then obtain the following expression for $\Sigma _{0}^{II}$:
\begin{eqnarray}
\Sigma _{0}^{II} & = &
 - \frac {1}{6r_{0}^{2} f_{1}} \left[ C + \ln \left( \frac {1}{r_{0}} {\sqrt {\frac {x}{f_{1}}}} \right) \right]
 \nonumber \\ & &
  - \frac {8}{r_{0}^{2}f_{1}} \Re \left\{ \int _{0}^{\infty } d\lambda
  \frac {\lambda }{1+ e^{2\pi \lambda }} \ln B \right\}
\nonumber \\ & &
  + O\left[x\ln \left( \frac {x}{r_{0}} \right) \right].
  %\nonumber \\
\label{eq:Sigma0twoI}
\end{eqnarray}
The integral in (\ref{eq:Sigma0twoI}) involving $\ln B$ cannot be computed exactly because we do not have
a closed-form expression for $B$ for all values of $\ell $.
However, this integral does not depend on $x$, and is finite because the behaviour of $B$ for large
$\lambda $ is given by (\ref{eq:BlargeL}).
Therefore we have, near the horizon,
\begin{equation}
\Sigma _{0}^{II} = - \frac {1}{12r_{0}^{2} f_{1}} \ln \left( \frac {x}{r_{0}} \right)  + {\mbox {finite terms}}.
\label{eq:Sigma02}
\end{equation}

Next we consider $\Sigma _{0}^{I}$ (\ref{eq:Sigma0one}).
From (\ref{eq:B}) with $\ell = \lambda - \frac {1}{2}$, we have
\begin{equation}
2B \frac {dB}{d\lambda } = 2\lambda + \frac {1}{3} r_{0}^{2} f_{1} \frac {dy_{1}}{d\lambda },
\end{equation}
so that, for large $\lambda $,
\begin{equation}
2B \frac {dB}{d\lambda } = 2\lambda + O(\lambda ^{-2}).
\end{equation}
Defining $B_{0} = B(\lambda = 0)$, and using the expansions (\ref{eq:GOhor}), the leading order behaviour
of $\Sigma _{0}^{I}$ can be written as
\begin{equation}
\Sigma _{0}^{I} = {\tilde {\Sigma }}_{1} + {\tilde {\Sigma }}_{2}  + O(x\ln [x/r_{0}]),
\label{eq:Sigma01}
\end{equation}
where
\begin{eqnarray}
{\tilde {\Sigma }}_{1} & = &
\int _{B_{0}}^{\infty } dB \left\{  \frac {4B}{r_{0}^{2} f_{1}}
 I_{0}\left( \frac {2B}{r_{0}} {\sqrt {\frac {x}{f_{1}}}} \right)
 K_{0}\left( \frac {2B}{r_{0}} {\sqrt {\frac {x}{f_{1}}}} \right)
 \right. \nonumber \\ & & \left.
  - \frac {d\lambda }{dB} \frac {1}{r_{0} {\sqrt {f_{1}x}}} \right\} ;
  \nonumber \\
{\tilde {\Sigma }}_{2} & = &
 \int _{0}^{\infty } d\lambda \left\{
\frac {4}{r_{0}^{2} f_{1}}
 I_{0}\left( \frac {2B}{r_{0}} {\sqrt {\frac {x}{f_{1}}}} \right)
 K_{0}\left( \frac {2B}{r_{0}} {\sqrt {\frac {x}{f_{1}}}} \right)
 \right. \nonumber \\ & & \left. \times
\left[ \lambda - B \frac {dB}{d\lambda } \right]
 \right\} .
\end{eqnarray}
To compute ${\tilde {\Sigma }}_{1}$, we first change the upper limit of the integral from $\lambda = \infty $
to $\lambda = \lambda _{L} \gg 1$, which simplifies the analysis.
The integral ${\tilde {\Sigma }}_{1}$ is in fact regular as $\lambda _{L} \rightarrow \infty $, but the individual
terms in it are not.
We use the standard result \cite{tomimatsu}, valid for all non-zero $\nu $:
\begin{eqnarray}
\int 2B I_{0}(B\nu ) K_{0}(B\nu ) \, dB & = &
B^{2} \left[ I_{0}(B\nu ) K_{0} (B\nu )
\right. \nonumber \\ & & \left.
+ I_{1} (B\nu ) K_{1}(B\nu ) \right] ,
\nonumber \\
\end{eqnarray}
which gives
\begin{eqnarray}
{\tilde {\Sigma }}_{1} & = &
\left[ \frac {2}{r_{0}^{2}f_{1}}  B^{2} \left\{
 I_{0}\left( \frac {2B}{r_{0}} {\sqrt {\frac {x}{f_{1}}}} \right)
 K_{0}\left( \frac {2B}{r_{0}} {\sqrt {\frac {x}{f_{1}}}} \right)
 \right. \right. \nonumber \\ & & \left. \left.
+  I_{1}\left( \frac {2B}{r_{0}} {\sqrt {\frac {x}{f_{1}}}} \right)
 K_{1}\left( \frac {2B}{r_{0}} {\sqrt {\frac {x}{f_{1}}}} \right)
 \right\}
 \right. \nonumber \\  & & \left.
 - \frac {\lambda (B)}{r_{0} {\sqrt {f_{1}x}}}
\right] _{B_{0}}^{B_{L}} ,
\label{eq:Sigma1one}
\end{eqnarray}
where $B_{L} = B(\lambda _{L})$.
For large $z$, we then use the expansion \cite{abramowitz}:
\begin{equation}
I_{n}(z) K_{n}(z) = \frac {1}{2z} \left[ 1 - \frac {4n^{2} -1}{8z^{2}} + O(z^{-4})
\right] ,
\end{equation}
together with the behaviour of $B$ for large $\ell $ (\ref{eq:BlargeL}).
This enables us to simplify the expression (\ref{eq:Sigma1one}) to obtain a quantity which is manifestly
finite as $\lambda _{L} \rightarrow \infty $, and then we can take the limit $\lambda _{L} \rightarrow \infty $.
The answer obtained is
\begin{eqnarray}
{\tilde {\Sigma }}_{1} & = &
-\frac {2B_{0}^{2}}{r_{0}^{2}f_{1}} \left[
 I_{0}\left( \frac {2B_{0}}{r_{0}} {\sqrt {\frac {x}{f_{1}}}} \right)
 K_{0}\left( \frac {2B_{0}}{r_{0}} {\sqrt {\frac {x}{f_{1}}}} \right)
 \right. \nonumber \\ & & \left.
+  I_{1}\left( \frac {2B_{0}}{r_{0}} {\sqrt {\frac {x}{f_{1}}}} \right)
 K_{1}\left( \frac {2B_{0}}{r_{0}} {\sqrt {\frac {x}{f_{1}}}} \right)
  \right] .
\end{eqnarray}
For small $x$, we now expand ${\tilde {\Sigma }}_{1}$, using (\ref{eq:I0K0hor}) and \cite{abramowitz}
\begin{equation}
I_{1}(x) K_{1}(x) = \frac {1}{2} + O(x^{2} \ln [x/r_{0}]) ,
\end{equation}
to give, for small $x$,
\begin{equation}
{\tilde {\Sigma }}_{1} = \frac {B_{0}^{2}}{r_{0}^{2}f_{1}} \ln \left( \frac {x}{r_{0}} \right)
+ {\mbox {finite terms}}.
\label{eq:Sigma1final}
\end{equation}
For ${\tilde {\Sigma }}_{2}$, it is sufficient to use the expansions (\ref{eq:I0K0hor})
in the integrand to give, for small $x$,
\begin{eqnarray}
{\tilde {\Sigma }}_{2} & = & -\frac {2}{r_{0}^{2}f_{1}} \ln \left( \frac {x}{r_{0}} \right)  \int _{0}^{\infty }
d\lambda \left( \lambda - B \frac {dB}{d\lambda } \right)
\nonumber \\ & &
+ {\mbox {finite terms}}
\nonumber \\ & = &
-\frac {1}{r_{0}^{2}f_{1}} \left[
B_{0}^{2} - \frac {1}{12} -  m^{2}r_{0}^{2}
- \left( \xi - \frac {1}{6} \right) R_{0}r_{0}^{2}
\right]
\nonumber \\ & &
\times \ln \left( \frac {x}{r_{0}} \right)
%\nonumber \\ & &
+ {\mbox {finite terms}}.
\label{eq:Sigma2final}
\end{eqnarray}
Combining (\ref{eq:Sigma1final}) and (\ref{eq:Sigma2final}), we find the following expression
for $\Sigma _{0}^{I}$ (\ref{eq:Sigma01}):
\begin{eqnarray}
\Sigma _{0}^{I} & = &
\frac {1}{r_{0}^{2}f_{1}}   \left[  \frac {1}{12} +  m^{2}r_{0}^{2}
+ \left( \xi - \frac {1}{6} \right) R_{0}r_{0}^{2}
\right] \ln \left( \frac {x}{r_{0}} \right)
\nonumber \\ & &
+ {\mbox {finite terms}},
\label{eq:Sigma01final}
\end{eqnarray}
and, combining this with (\ref{eq:Sigma02}), gives our final expression for the behaviour of the mode sum
$\Sigma $ (\ref{eq:modesum}) near the horizon:
\begin{eqnarray}
\Sigma  & = &
\frac {T}{4\pi f_{1}}   \left[  m^{2}
+ \left( \xi - \frac {1}{6} \right) R_{0}
\right] \ln \left( \frac {x}{r_{0}} \right)
+ {\mbox {finite terms}}
\nonumber \\ & = &
\frac {1}{16\pi ^{2}} \left[  m^{2}
+ \left( \xi - \frac {1}{6} \right) R_{0}
\right] \ln \left(  \frac {x}{r_{0}} \right)
+ {\mbox {finite terms}},
\nonumber \\
\label{eq:Sigmafinal}
\end{eqnarray}
where we have used the fact that the temperature $T=\kappa /2\pi $, where $\kappa =f_{1}/2$ is the surface
gravity of the horizon.

From (\ref{eq:Sigmafinal}) we see that the mode sum $\Sigma $ diverges near a horizon unless the quantum
scalar field is massless and conformally coupled.
This behaviour can be seen in our numerical results in Figs. \ref{fig:phi2numm0}--\ref{fig:phi2numm1close},
where it is apparent that $\Sigma $ is regular at a horizon in the massless, conformally coupled,
case but otherwise divergent.

Finally, comparing (\ref{eq:phi2analyticII}) and (\ref{eq:Sigmafinal}), we see that the logarithmic divergences
cancel and $\Sigma +  \langle \phi ^{2} \rangle _{\rm {analytic}}^{II}$ is regular at the horizon.
Since we have already shown that the remaining contributions to $\langle \phi ^{2} \rangle _{\rm {ren}}$,
namely $\langle \phi ^{2} \rangle _{\rm {analytic}}^{I}$ (\ref{eq:phi2analyticII}) and
$\langle \phi ^{2} \rangle _{\rm {numeric}}^{II}$ (\ref{eq:phi2numericII}), are regular at a horizon,
we therefore conclude that the total $\langle \phi ^{2} \rangle _{\rm {ren}}$, for a thermal
state at the same temperature as the horizon, is regular at that horizon.
For the particular case of lukewarm black holes, the analysis above applies equally well to the event
and cosmological horizons, which are at the same temperature.
Therefore, for a thermal state at this temperature, we have shown analytically that
$\langle \phi ^{2} \rangle _{\rm {ren}}$ is finite at both the event and cosmological horizons.
This means that the divergences seen  (Fig. \ref{fig:phi2renm1}) in
$\langle \phi ^{2} \rangle _{\rm {analytic}}$ and $\langle \phi ^{2} \rangle _{\rm {numeric}}$
when the scalar field is no longer massless do in fact cancel to give a finite
$\langle \phi ^{2} \rangle _{\rm {ren}}$.

\section{Conclusions}
\label{sec:conc}

In this paper we have studied the renormalized expectation value $\langle \phi ^{2} \rangle _{\rm {ren}}$
for a quantum scalar field on a lukewarm black hole background, where the space-time possesses a black hole
event horizon and a cosmological horizon which are at the same temperature.
We have used the method of \cite{ash} to compute $\langle \phi ^{2} \rangle _{\rm {ren}}$
for a thermal quantum state at this natural temperature.
Our numerical computations have indicated that $\langle \phi ^{2} \rangle _{\rm {ren}}$ is regular
on both the event and cosmological horizons, and this has been proved analytically as well.

The method of Ref. \cite{ash} splits $\langle \phi ^{2} \rangle _{\rm {ren}}$ into two parts,
an analytic expression $\langle \phi ^{2} \rangle _{\rm {analytic}}$ and a part,
$\langle \phi ^{2} \rangle _{\rm {numeric}}$, which can only be computed numerically.
One might hope that the analytic expression would be a good approximation to the exact quantity,
at least in some limit (say, for large mass).
However, as observed in Ref. \cite{ash}, this is not the case because $\langle \phi ^{2} \rangle _{\rm {analytic}}$
contains a term which diverges logarithmically near a horizon unless the quantum scalar field is massless and
conformally coupled.
We have shown that, for a thermal state at the same temperature as the horizon,
this divergence cancels with a divergence in
$\langle \phi ^{2} \rangle _{\rm {numeric}}$ near the horizon to give an overall finite quantity.
In the literature other analytic approximations have been developed \cite{phi2approx,Tmunuapprox},
some of which do not have this logarithmic divergence.

Unlike the Schwarzschild-de Sitter black hole \cite{sds}, for lukewarm Reissner-Nordstr\"om-de Sitter
black holes we have shown that a static quantum state can be constructed which
has a regular renormalized expectation value $\langle \phi ^{2} \rangle _{\rm {ren}}$  across both future and past
event and cosmological horizons.
This is the equivalent of the Hartle-Hawking state \cite{hh} for these black holes.
This result is in accordance with the theorems of Kay and Wald \cite{kay}, who proved that no regular thermal
state can exist on a Reissner-Nordstr\"om-de Sitter black hole for which the temperatures of the event and
cosmological horizons are not equal.
When the temperatures do match, the fact that the region between the event and cosmological horizons
has a regular Euclidean section \cite{mellor} allows one to construct a thermal state
at the natural temperature and then we have shown that this state has regular
$\langle \phi ^{2} \rangle _{\rm {ren}}$ on both the event and cosmological horizons.

Our work in this paper has considered only $\langle \phi ^{2} \rangle _{\rm {ren}}$, and not the RSET.
The computation of the RSET, following \cite{ash}, mirrors that here for $\langle \phi ^{2} \rangle _{\rm {ren}}$,
but is considerably more complex.
In particular, even though we have shown that $\langle \phi ^{2} \rangle _{\rm {ren}}$ is finite
on both the event and cosmological horizons, this does not guarantee that the RSET will be finite there.
We plan to report on this in the near future.

%\bigskip

\begin{acknowledgments}
We thank Adrian Ottewill for numerous invaluable discussions and insights into the calculations
involved in this paper.
We also thank Giles Martin and Calvin Smith for useful discussions, particularly on Hadamard series,
and Bernard Kay for helpful insight.
The work of EW is supported by PPARC UK, grant reference number PPA/G/S/2003/00082,
while that of PMY is supported by EPSRC UK.
\end{acknowledgments}

\end{document}